# Underdoped cuprates, manifestations of boson-fermion crossover, 'quantum oscillations' and the robust fractional quantum Hall state $\nu = {}^2/_5$

John A. Wilson

H. H. Wills Physics Laboratory
University of Bristol
Bristol BS8 1TL,  U.K.

**Abstract**

The quantum oscillation data accruing from a whole variety of underdoped HTSC systems and physical measurements has almost universally been presented in very classical terms following the Lifshitz-Kosevich formulation for Fermi-Landau quasiparticles, quantized into vortices and Landau levels under the high applied magnetic fields.  The Fermi surface, as allegedly monitored, always emerges as being a very small fraction of the large parent Fermi surface detected at higher doping.  This then calls for some Fermi surface reconstruction to have occurred, as from a charged density wave, etc. etc..  The present paper refutes this reciprocal space approach for the underdoped HTSC cuprates, and shows that many of the more recent detailed works taking this line are not consistent with the data.  The present author introduced an alternative real-space interpretation of these matters a couple of years ago.  This was in terms of interaction there of the vortex array with the stripe array, using a 2D modelling of the latter.  That approach has in the present paper been revised somewhat and copes well with addressing the wide variety of new data now available.  The paper pursues and justifies the BEC/BCS, two-subsystem, negative-$U$ based modelling advocated earlier for the HTSC cuprates, along with the strength and geometry of the striping acquired in a strong magnetic field.  A wide variety of recent experiments are expounded in these terms to support an overall appreciation of a situation far from being appropriately treated within a Fermi-Landau quasiparticle framework.  Indeed it is demonstrated that the universality of the 'QO' observations ensues from the highly correlated physics of the fractional quantum Hall effect. The data all point to FQHE state with filling factor $\nu = {}^2/_5$.  This particular state in its genesis and expression finds a much greater alignment with the BEC/BCS crossover condition and the proximity of the HTSC materials to Mott-Anderson localization than does any model for the prevailing inhomogeneous electronic conditions which appeals to well-metallized Fermi-Landau quasiparticle physics.

**Contents**

**§1. Introduction to area of concern.**

The consequences of boson-fermion (B-F) crossover in relation to high temperature superconductivity (HTSC) have been probed theoretically for a considerable time now [1-6]. The mixed-valent (II/III - "hole-doped") superconducting cuprates have over a yet longer period been presented by the current author as embracing such a situation, established there within a negative-$U$ setting [7-10 and refs. therein]. In the past two or three years attention regarding these electronically inhomogeneous HTSC systems has been drawn back into re-examination of the basic transport properties, both as functions of temperature and magnetic field, with the field variable now being extended up to very high values [11-13]. This fresh attention follows striking revelations made initially via AMRO and subsequently via dHS/dHVA quantum oscillation techniques, first on samples of high hole content, $p$, and then duly working back into the underdoped regime [14-19]. While strongly overdoped material exhibits evidence of fairly conventional if clearly correlated Fermi liquid behaviour, increasingly strong departure from such behaviour is met with for $p$ values below ~0.25. It has accordingly been particularly disconcerting to see a great many recent publications look toward rather conventional treatments of the Fermiology in the HTSC cuprate systems in samples for which $p$ is ~ 0.13 or less. Such works, both experimental and theoretical, have sought to account for the data gathered there simply in terms of band-folding and conventional Fermiology, entertaining rather standard notions of SDW, CDW or stripe phase formation for what is underway [20-25]. In my earlier paper on this matter [27] I proposed a radically different real space scenario of how such an oscillatory signal might arise. This involves the ratcheted passage of flux vortices between field-enhanced stripe domains. The present paper marks a certain revision of that appreciation that focusses now not so much on the stripe domain's size as upon its hole content. A detailed argument is developed of how the underdoped samples with $p \approx 0.10$ do not support Fermi liquid behaviour, but are instead representatives of the Fractional Quantum Hall condition (FQHE) $\nu = {}^2/_5$. The latter is one highly appropriate to the resonant Feshbach state long advocated by the author as origin of the HTSC phenomena [7-10, 26-28].

Section 3 in the present work emphasizes the way in which the standard electrical transport behaviour stands reinterpretion in terms of coexisting fermions (coherent and incoherent) and bosons. Section 4 endeavours to supply a more detailed appreciation of the stripe state condition developing in these underdoped HTSC materials in the pseudogap regime. Section 5 looks at the enhancement of that stripe condition secured under application of a strong magnetic field. Section 6 reworks the real-space understanding of the 'quantum oscillation' data forthcoming from HTSC material with $p \sim 0.1$ and indicates how it is not compatible with a CDW/SDW Fermi Surface reconstruction scenario. Section 7 explores the appreciation gained in Section 6 that all these results are a consequence of the adoption of the FQHE $\nu = {}^2/_5$ state, and relates this state of affairs to the B-F crossover event within the negative-$U$ environment afforded by the mixed-valent cuprates.

**§2. Background to Boson-Fermion (B-F) crossover modelling of HTSC cuprates.**

Already I have laid out in very considerable detail how I advocate interpretation of all HTSC cuprate data to proceed, whether issuing from photoemission and scanning tunnelling microscopy [see 26], transport and quantum oscillations [see 27], or pump/probe and optical experiments [see 28]. Throughout this interpretative work a critical concentration of $p_c$ = 0.185, flagged up a dozen years ago now within the thermodynamic studies by Loram *et al.* [29], stands as the hole concentration at which the superconducting condensation energy per Cu atom comes to a sharply cusped maximum. It constitutes the 'hole doping' level below which the quasiparticles, starting with those on the band structural saddles, incur a significant incoherence even at low temperatures. In the B-F modelling this incoherence is presented as the consequence of a chronic scattering in mixed-valent HTSC cuprates associated with the degeneracy in play there between the primary quasiparticles and the negative-$U$ instigated local-pair bosons. This strong scattering takes on two forms: (1) the *e*-on-*e*-to-*b* scattering of boson creation, origin of the strong $T^2$ contribution to $\rho(T)$ witnessed over a wide range of $p$ to either side of $p_c$, and (2) the *h*-on-(uncondensed)-*b* scattering, source of a strong and extensive *T*-linear contribution to $\rho(T)$ in evidence down to notably low *T* [27]. By $T_c(p_c)$ the above two strong scattering processes have become of very comparable intensity [12]. The critical concentration $p_c$ stands as that at which the production and maintenance of active bosons is most favoured. In addition to supplying maximum condensation energy per Cu atom, $\epsilon$, to the overall condensate, $p_c$ is the composition upholding maximum superfluid number density, $n_s$, as has been evidenced from μSR penetration depth measurements [30]. Maximum $T_c$ arrives, by contrast, at the slightly reduced $p$ value of 0.16. Below this latter composition $T_c$ declines linearly with $n_s$. The condensation energy on the other hand falls away much more sharply than does $T_c(p)$ [29], and right from $p_c$, as the instigating negative-$U$ state is abstracted to greater binding energy, $U(p)$. The quasiparticles of the Fermi surface, beginning with those in the saddle regions, individually start to de-cohere and gap under the prevailing chronic *e-e* and *e-b* scattering. The associated move towards Mott-Anderson localization fosters the general loss of coherence and introduces additional strong scattering as *T*, this with falling *p* progressively extending up to very high temperatures, as first was reported back in 1987 by Gurvitch and Fiory [31]. With the reduction in *p* and the growing pseudogapping to the quasiparticle DOS as local-pair binding energy, $U(p)$, and charge trapping energy, $£(p)$, each augment, the superconducting gap parameter, $\Delta(p)$, itself retreats steadily from its maximum – this the so-called 'gapping dichotomy', much in evidence in a wide variety of experiments on underdoped HTSC samples [26-28].

With the above perception as to what is driving pseudogapping in the underdoped (sub $p$ = 0.185) regime, it seems quite inconceivable how standard Fermiology might be deemed appropriate to account for the recently extracted high-field quantum oscillation results from low-*p* samples. In [27] I previously have set out how such low temperature data from underdoped HTSC samples would support interpretation as coming *not* from standard, maximal orbit, *k-space*-based periodicities in 1/*B* (after the Lifshitz-Kosevich formulation elaborated for conventional metals) *but instead* from a quantized *real*-space sourcing.

The latter scenario incorporates a 2D stripe-phase appreciation [32,33] of the situation prevailing in the underdoped mixed-valent cuprates under charge segregation and superlattice formation prior even to any application of a magnetic field. In the striping process the 45°-oriented square domains there become filled by divalent cuprate coordination units ($^9$Cu$_{II}^0$), with the latter domains bounded by walls holding alternately spaced hole-bearing coordination units ($^8$Cu$_{III}^0$) [8,32]. The negative-*U* local pairs ($^{10}$Cu$_{III}^{2-}$) materializing at appropriately sited Cu$_{III}$ coordination units are free then to propagate over the mixed-valent peripheries of the domains. By contrast the spins and charges within their divalent inner regions become progressively gapped and confined as *p* and/or *T* are reduced and one approaches Mott-Anderson localization. For as long as the spins prefer local RVB spin-pairing and avoid LRO antiferromagnetic order, the coupled ensemble of fermionic pairs and local pair bosons avoids strong spin-flip pair-breaking within the globally permeating superconducting condensate. Note that while the pseudogap energy varies wildly with precise location, STM discloses that the superconducting (inner) gap is effectively homogeneous [34, fig.2a]. At all temperatures below $T_c$ some proportion both of the local-pair bosons and of the fermionic quasiparticles continue, nonetheless, to exist outside the condensate. Local pairs in fact remain able to form in the immediate vicinity of the negative-*U* centres on the domain boundaries up to well beyond $T_c$ – and indeed beyond the Nernst regime and even *T\**. They transiently are generated up to the highest temperatures, and ultimately one witnesses $\rho$ in these underdoped samples taken to values above the Mott-Ioffe-Regel limit (i.e. into full incoherence).

Note in the real space understanding of the quantum oscillation results advanced in [27], the 1/*B* oscillations (realized in applied fields $H_{app}$>$H_{irrev}$) have been proposed to arise as the number of 'stripe' domains becomes shed, one by one in ratchet fashion, from the footprint, *F*(*H*), allotted to each flux vortex at any specific $H_{app}$, within a still quantized magnetic flux array. Prior to turning to develop this model and the wealth of new information in this area continuing to accumulate from STM, ARPES and QO work, it first is advantageous to examine how the details of the very recent high-field measurements of $\rho$ and $R_H$ reported by Hussey and coauthors in [13] are to be accommodated within the above scenario.

### §3. Critique of the application of standard electron-hole modelling to low *T*, high *H* transport properties of underdoped HTSC cuprates.

As is divulged by the steady conversion of the Hall and Seebeck coefficients at low temperatures from 'hole' towards 'electron' behaviour observed universally with underdoped HTSC cuprates under high magnetic fields (typically > 30 tesla), some two-component approach would seem called for. What those two components might be then becomes the question. As is noted above much modelling of the high-field quantum oscillation (QO) results has to date revolved around band-folding under the presumed impress of rather conventional

SDW or CDW/PLD formation. A prime problem with all such interpretations is that the *n*-type, small area segments of Fermi surface seemingly implicated by the QO data would then issue precisely from those parts of the original large Fermi surface wherein the quasiparticle states undergo their greatest scattering and loss of coherence. Hence why such states should go forward to dominate the high-field transport is highly questionable. Despite this fundamental objection, raised previously in [27], the examination and interpretation offered in [13] has, as elsewhere, proceeded in this vein.

When one takes the two components involved in sourcing the data of [13] to be standardly reconfigured hole and electron quasiparticles, one is drawn to make application of the quasi-classical equations for the net longitudinal and transverse (Hall) resistivities

$$\rho_{xx}(H) = \frac{(\sigma_h + \sigma_e) + \sigma_h \sigma_e (\sigma_h R_h^2 + \sigma_e R_e^2).H^2}{(\sigma_h + \sigma_e)^2 + \sigma_h^2 \sigma_e^2 (R_h - R_e)^2.H^2} \quad ,$$

$$\rho_{xy}(H) = \frac{(\sigma_h^2 R_h - \sigma_e^2 R_e) - \sigma_h^2 \sigma_e^2 R_h R_e (R_h - R_e).H^2}{(\sigma_h + \sigma_e)^2 + \sigma_h^2 \sigma_e^2 (R_h - R_e)^2.H^2} .H \quad .$$

In order to render analysis of the high-field data along these lines tractable, one has to presume that the numbers of particles involved here are field-independent, as too their mobilities. Moreover, as was addressed in [13], one must be aware of the possible complications inserted by the chain carriers for the material specifically under investigation here - namely $YBa_2Cu_4O_8$ ($T_{c,H=0}$ = 80 K, $p \approx {}^1/_8$), – featuring too in the early QO work of [16a,b]. The Hall data of [13] have been obtained with $H//c$, $I//a$, and $V_H//b$ (*b* being the chain direction). Experimentally one finds that beyond ~40 tesla (where the system passes, it is being claimed, into the 'normal state' condition) both $\rho_{xx}$ and $\rho_{xy}$ become essentially straight-line functions of *H*, with the *electron* conductivity (within this standard approach) becoming dominant here at low temperatures over the hole conductivity. The full $\rho_{xy}$ data set collected in [13] is reproduced in **figure 1**. **Figure 2** presents associated $\rho_{xx}$ and $\rho_{xy}$ data for three selected temperatures, together with fittings made in the above high-field region, plus their extrapolation back into the ordinarily superconducting low-field region. All fittings are executed *simultaneously* for both $\rho$ and $R_H$ in keeping with the above pair of equations. From this overall view the authors of reference [13] proceed then to extract as functions of temperature the two individual component carrier contributions that best are able to provide the above levels of fitting to the *high*-field data $\rho_{xx,xy}(T,H')$ in the straight segments $H'$ > ~40 T. Before examining how they attempt to account for the outcome to their two-component separation accomplished in [13,figure 3], here reproduced in the present **figure 3** with slight modification, let us first take a look at how the boson-fermion crossover model can cope with supplying for this data a seemingly more appropriate two-component description.

For the moment let us set aside the matter of stripe formation (be this diagonal-square or otherwise) and here concentrate upon the parent state carriers. Firstly we have the fermions of the heavily pseudogapped large Fermi surface, hole-like in character wherever and whenever relating still to coherent quasiparticles encompassing the '$\pi,\pi$' location. Secondly we have that subset of local-pair and induced bosons which, within the crossover modelling, still reside outside any bosonic condensate under the given conditions of $T$, $H$ and the specific level of underdoping $p'$ (here $\sim 1/8$). We accordingly shall associate $\sigma_h$ and $R_h$ in the underdoped material with the residual coherent fermion content, and $\sigma_e$ and $R_e$ with the above bosons together with any incoherent fermions. The bosons, being not subject to the state single-occupancy constraint of Fermi statistics, will of course always respond as negatively charged carriers. Inspection of figure 3 would from this viewpoint (a figure that recall is anchored in the large-$H$ regime of figure 1) indicate that up beyond 100 K the bosonic contribution, $\sigma_e$, to the overall *conductivity*, $\sigma_{xx}$, is tending to zero. The mobility/coherency of maintained bosons is by that stage becoming much reduced. In due course the hole-like fermions come to dominate $\sigma$, despite their own contribution $\sigma_h$ to $\sigma_{xx}$ itself dropping off as $1/(\rho_o + AT^2)$, a outcome of the strong fermion-on-fermion scattering proximate to the negative-$U$ centres and the transient boson generation. $R_h$, the contribution of the coherent fermions to the overall Hall coefficient $R_H$, becomes, once above 30 K and at these high fields, more or less independent of temperature. Upon passing *down* through this same temperature range from 100 K to 30 K, $\sigma_e$, the bosonic contribution to $\sigma_{xx}$, climbs approximately as $1/T$ as a result of the reduction in phase space open to boson-fermion scattering. For the given $p'$, the actual number of *free* bosons and incoherent electrons across the above temperature range remains quite appreciable and accordingly in this material these will make rather limited contribution to $R_H$.

Once below 30 K it becomes very apparent from figure 3 that a totally new regime of behaviour is being entered into. The fermionic hole component starts to register exponential removal. This comes from two sources. Firstly many more quasiparticles are experiencing conversion to bosonic local pairs ($^{10}Cu_{III}^{2-}$). Secondly at these low temperatures more and more of the remaining fermions become trapped polaronically into weak localization. The observed rather sharp commencement could well be coupled to the onset of pinned 2D striping within the prevailing conditions of low $T$ and high $H$ (at the current $p' \sim 1/8$). Not only does $\sigma_h$ diminish exponentially, but, in the first instance, $R_h$ pointedly veers strongly upwards. While such changes are occurring to the fermionic hole content, what of the bosonic behaviour? The bosonic contribution $\sigma_e$, growing hitherto as $1/T$, now plateaus out below 20 K. Possibly stripe pinning to contiguous dopant centres and a growing confinement of the bosons to and within the stripes stand responsible for such capping of $\sigma_e$ at low $T$. An additional matter is whether the onset of stripe and carrier pinning and trapping becomes advanced in the present ($T,H,p'$) conditions by the field vortices themselves settling into a regular *square* array, responding to the geometry of the square stripe domain array. One sees $R_e$ begin at the same juncture below 35 K (and $H$>35 T) to move toward larger values, as if the active bosons are being abstracted into the superconducting condensate. Ultimately $R_e$, as with $\sigma_e$, levels out towards helium temperatures, implying that a capped residue of bosons to remain both untrapped and outside the condensate. Under these low temperature conditions, $R_h$, by now largely deriving from the residual near-nodal carriers, wheels around towards zero in superconductive fashion – the global condition evident in figure 2a for $T$=20 K once $H$<25 tesla.

**§4. Recent experimental results bearing on the form and nature of charge striping.**

*4.1. Background to striping, with special reference to incommensurate 1T and 2H $TaS_2/Se_2$.*

Since in the above, and again in what follows, the concept of organized structuring of the inhomogeneity inherent in the 'doped' mixed-valent HTSC systems pervades the discussion, we must examine how the latest research embellishes understanding of the 'striped' construction, be this one-dimensional, as most have adhered to [35], or two-dimensional, as I have long advanced [32,33] (see also a Ginzburg-Landau study to appear from Melikyan & Norman). Around this latter distinction lies, I believe, the key to understanding very many of the puzzles presented by the underdoped cuprate materials and especially now the quantum oscillation data [15-19].

Many speak of the above 'incommensurate' charge and spin ordering apparent in elastic and especially inelastic neutron [36] and X-ray scattering [37] experiments in terms of charge and spin density wave formation, as if dictated by the Fermi surface geometry and the 'nesting' thereof. However three points call for recognition here: (*1*) the essentially circular Fermi surface lacks strong nesting characteristics; (*2*) the incoherency of the Fermi sea, i.e. lack of sharpness to the Fermi surface, particularly within its potentially better nested segments, the saddles; (*3*) the appearance of comparably 'incommensurate' diffraction from the related *non*-metallic mixed-valent nickelate systems [38], addressed earlier in [32,§3.3]. The latter results point to the most likely origin of the attempted structural organization in the mixed-valent cuprates being lattice strain. This develops when, at the local level, the charge inhomogeneity finds itself expressed as differences in chemical bonding within the differently charge-loaded coordination units. The structural differences imposed here by the $d_z^2$ and $d_{x^2-y^2}$ states of the d($e_g$)p-$\sigma\sigma^*$ set upon the on-coordination-unit-based, Cu-O interaction become very marked. They require to be addressed on a local basis [32(§3.3)] in terms of the Jahn-Teller effect, active within our current systems/structures at $d^7$ and $d^9$. The lattice response to the local interaction is evidenced in statistical fashion by the EXAFS-monitored variability in nearest-neighbour Cu-O bond length [39], and still more directly so in STM-acquired scans of the lattice array which display highly local electronic characteristics [40]. The pseudogap behaviour shows much greater sensitivity here to local circumstances than does the more collective superconductive 'inner' gapping of low temperatures.

What structurally orders, or attempts to, within the cuprates is not the actual location of the dopant/substituent atoms themselves but the deficit charge content that issues from the latter, in conjunction with concomitant organization of the spin array. Because the order is primarily a product of the outer (conduction) electrons, the associated diffraction data become more evanescent and of a lower magnitude than in the case of structural ordering in an alloy such as Cu/Au or $Cu_{10}Sb_3$, in which the atomic volume/mass differences between ordering components are very marked. $Cu_{10}Sb_3$, as it happens, yields an identical diffraction pattern to layer-structured $1T_3$-$TaS_2$ – the subscript 3 here denotes the low temperature, 3-by-1 rotated, $\sqrt{13}a_o$, hexagonal (bar stacking) superlattice condition [41,42]. This 'commensurate' order of $1T_3$-$TaS_2$ follows higher temperature states designated $1T_2$ and $1T_1$, strongly 'discommensurate' and 'incommensurate' respectively. The latter phase in this $5d^1$ delocalized metal demonstrably does relate back to the Fermi surface, and accordingly warrants treatment in terms primarily of Fermi surface nesting and CDW formation. However, even with $TaS_2$, the $1T_2$ state reveals in its temperature evolution a strong participation of the lattice in the electronic processes. The 'discommensurate' nature of $1T_2$-$TaS_2$ is observed to be one of successive domains of 'locked-in' lattice commensuration (based upon rotated 13-atom clustering) set around by phase-slip boundaries [43,44]. The rather narrow width of those boundaries is a marker of the continued drive in $1T_2$ to cluster formation. At higher temperatures however, as was intimated above, the superstate wavevector within *un*rotated $1T_1$-$TaS_2$ can via Ti substitution be forced away from such commensuration and then manifestly does reflect the gradually changing dimensions of the Fermi surface, so justifying the CDW label. In the alternative 2H polytypic form of $TaS_2$ one again discovers a CDW/PLD state to arise at reduced temperature, now of suitably modified wavevector and appreciably lower amplitude. This time at low temperatures the 2H polytype shows commensurate lock-in to the lattice at $3a_o$, following what was at first perceived as a simple incommensurate condition [41]. However once more the onset phase proves discommensurate [45]. The diffraction pattern via its weak, higher-order, spotting manifests such a state of affairs - easy to miss or neglect. What proved highly fortuitous as regards experimental observation of this discommensurate phase in 2H-$TaSe_2$ was that the two-sandwich stacking of the 2H polytype confers hexagonal symmetry breakage upon the resulting CDW structures, commensurate regions becoming very slightly orthorhombic. This symmetry breakage brings to *dark-field* imaging electron microscopy strong *areal* brightness contrast between appropriate near-neighbour domains, within an overall 'double-hexagonal' domain geometry. I do not wish to proceed further into the details of this history other than to point to three matters of immediate relevance now to the HTSC situation: namely (*1*) areal vs linear diffraction contrast, (*2*) 2D vs 1D domain structuring, (*3*) the dynamics appertaining to the discommensurate condition.

*4.2. Areal versus linear diffraction contrast.*

In connection with point (*1*) it is evident a disappointingly small amount of low temperature, transmission electron microscopy has been undertaken on underdoped HTSC systems to probe these important questions. Clearly greater obstacles to such work exist here than was the case with layered, stoichiometric, hexagonal 2H-$TaSe_2$. Notwithstanding, BSCCO, particularly when underdoped, offers clear potential. Early TEM work on overdoped $Tl_2Ba_2CuO_{6+\delta}$ from Hewat *et al* [46] uncovered structural complexities which it would be extremely difficult for other techniques to pick up, involving microtwinning of a type quite distinct from the basic orthorhombic twinning. Detail recorded by the surface probes of STM on CNCOC and subsequently BSCCO, and termed there 'checker-boarding' [47], similarly emphasizes the HTSC materials to be structurally more complex at the micro-level than many are willing to embrace. From the observed wave-vectors it is apparent the 'checker-boarding' reported in [48] is not the charge/spin 'striping' probed in bulk diffraction experiments, whether using synchrotron X-rays or neutrons.

*4.3. 1q versus 2q discommensuration structuring.*

In connection with point (*2*) I never have been convinced that the latter diffraction techniques directly reveal the striping in the HTSC cuprates to be other than two-dimensional. Considering that virtually all the HTSC families are orthorhombic, a one-dimensional aspect to the incommensurate diffraction features in question might not have been overly surprising. Both 1T- and 2H-$TaS_2$ do in fact initially exhibit upon warming a 1**q** as distinct from multi-**q** ICDW variant [42]. The IC diffraction from under-doped as from optimally-doped YBCO-123 is however, despite the presence of the chains and even for detwinned crystals, always (pseudo-)tetragonal in disposition [49]: the important recent work from Hinkov *et al* in this regard will be addressed in detail below. Customarily with LBCO and derivatives it has been asserted that the observed 2D aspect to the overall diffraction pattern is the outcome simply of a 'crossed' projection from successive 1D layering in the multi-layered, low temperature structural phases (LTO and LTT) [35]. However $HgBa_2CuO_{4+\delta}$ [50] and $Bi_2Sr_2CuO_{6+\delta}$ do not have this complication to appeal to and yet show 2D symmetry in their diffraction.

When originally drawing up 2D patterning of seemingly appropriate stripe structures for the HTSC materials in [51,8] I adopted the 'natural' course of aligning those stripes in the *x* and *y* (saddle or antinodal) directions, this being in keeping with the detected axial orientation of the spot displacement vectors, δ, away from the primary (π,π) locations set by the 2-by-2 antiferromagnetic diffraction spotting of the undoped parents. In [51] I pointed out that this AF order entails a face-centred spin geometry which means many diffraction spots (e.g. 1,0) become forbidden for the primary structure and the same likewise will follow for the visibility of the derivative IC satellites. It soon was realized though that, as with the multi-**q** structuring of the discommensurations (dc's) in 2H-$TaSe_2$, it should be possible to have 45°-inclined, 'partial' phase-slip boundaries capable of generating the relevant diffraction, as set out earlier in [32]. Two such 'partial' phase-slip boundaries stand equivalent to one of the original type in bringing to $\delta$ = ⅛ material axial spin-repeat over an (approximately) 8$a_o$ lattice period, as detected experimentally (but see §6.2). Moreover this revised organization of the stripes holds the potential to account [32] for the Magnetic Circular Dichroism (MCD) data of Kaminski *et al* [52], without necessitating any recourse to the local current-loop model advocated by Varma and colleagues [53]. In the 45° stripe modelling the appropriate magnetic/time reversal symmetry breaking comes from compliance of the local spin orientation within the domains to the stripes themselves under magneto-elastic coupling. The latter action, it would appear, derives from strong Jahn-Teller coupling associated with the dominantly d$^9$ coordination unit loading ($^9$Cu$_{II}^0$) of the domain interiors. The resulting majority clockwise and anti-clockwise circulations of the locally AF spin alignments within successive 45°-oriented domains turns out to be precisely as has been deduced by Fine directly from the neutron diffraction pattern [54], and that within the presentation of [32] is a direct product of the hole charge ordering of the domain walls.

In figure 1 of [32] I illustrated the situation for $p = 1/8$ (when taken to conform to commensurate geometry $\delta = 1/8$), and this figure now is reproduced in **figure 4**, **al**though employing a slightly different mode of representation of the frozen charge and spin array. **Figure 5** makes analogous representation of the $\delta = 1/7$ case, here for charge loading $p = 8/7^2$ or 0.1632, i.e. just beyond $T_c^{max}(p)$. Note in 2D it is not possible to construct a regular pattern of the current type wherein $\delta = 1/7$ can be matched to hole content $p = 7/7^2$. Furthermore note that within the patterning of Figure 5 the geometry of hole accommodation around the rotated domain corners is of two distinct dispositions, only the entire $7a_o \times 7a_o$ square defining the full charge-ordered repeat. **Figure 6** portrays in like fashion the case for $9a_o$ geometry, where now the hole content at $p = 9/9^2$ does become regularly spaced, although once more exhibiting different hole dispositions around its cell-centre and cell-corner dc crossings.

An important point to register here is that despite the regular patterning of fig. 4 at $\delta = 1/8$ and $p = 8/8^2$ being geometrically very attractive, recent refined neutron work on LBCO [55] has affirmed what had been implicit in work on LSCO [56,51] – namely that, contrary to general acceptance, there is in fact for (La$_{2-x}$Ba$_X$)CuO$_4$ at $x = p = 1/8$ no full commensurate lock-in to $\delta = 1/8$: rather the experimentally measured $\delta$ value lies significantly below $1/8$. This is a matter for which solution was offerred in [51], and is one to which we shall return in due course in §6.2 in connection with the quantum oscillation results.

Before leaving the matter of 2D vs. 1D as regards true representation of the experimental diffraction information, attention finally is drawn to Azzouz and coworkers' [57] recently reworked analysis of the HTSC cuprate diffraction data in terms of spiral antiferromagnetism, this quite independent of my own [32] or Fine's [54] deliberations. As was intimated above, my understanding is that such an exclusively magnetic view is not in fact appropriate to the present situation. The same would go for regarding the magnetic array as an example of a Skyrmionic lattice [58] should one tie that term to its original sense of arising from the Dzyaloshinskii-Moriya magnetic interaction within a non-centrosymmetric host. The dominant physics in the HTSC case surely is not magnetic but derives from charge segregation within a mixed-valent system barely removed from the Mott-Anderson transition. The neutron-sensed *areal* magnetic ordering in underdoped HTSC materials largely is impressed by the linear, neutron-invisible, charge ordering and its associated lattice strain arising from the very strong J.T. effect. Anisotropic forces due to magneto-elastic coupling impose the chiral spin geometry. The domains are, remember, only ~25 Å across. One additionally needs to keep in mind here the S=½ nature of the Cu(II) spins, their strong *p-d* hybridization at the termination of the 3*d* (1$^{st}$ T.M.) series, their RVB potential, and the invariably frustrated form to their 3D coupling, to perceive why overplaying consideration of standard magnetism and downplaying its spin-gapping is dangerous. A spin gap of 10 meV is, recall, equivalent to 120 K. The development of site charge differentiation stands paramount in governing the nature of the pseudogap condition from which HTSC springs.

*4.4. The dynamics of stripe behaviour: the electrical noise studies of Caplan et al. [59].*

We now come to point (*3*) regarding the stability of the stripe array. It is clear from the diffraction information that for most HTSC materials (away from low-*p* LBCO and its RE-substituted derivatives) striping is not *ordinarily* frozen in, temporally or spatially, to procure stable, long-range structuring. The diffraction data, nevertheless, from spot widths do make evident that quite appreciable correlation times and lengths are building. A very revealing additional measure of the meso-scale order developing quite sharply once below ~ 270 K is to be gained from examination of the associated *extra* electrical noise experienced there at very low frequencies.

Caplan *et al* [59] have made a close study of this electrical noise – in particular of the excitation energies involved, of its magnetic sensitivity, of matters such as aging and hysteresis, and above all of the "large-fluctuator" characteristics that appertain to the stabilizing stripe condition. These authors have monitored the 1/*f* noise over 8 octaves from 0.3 to 112 Hz, presenting their data in octave-compressed bands. Appropriately low thermal rates of change (~0.2 K/min) were used. The clarity of detail reported in the paper derives from the tiny volume and refined nature of their samples. The films are grown on $SrTiO_3$ and $LaAlO_3$ by laser deposition, are just 30 nm thick, and are photo-lithographically configured and ion-milled to be ~2 to 3 μm wide and 15 μm long. These films then have been post-annealed in low pressure $O_2$ to obtain the desired levels of underdoping ($T_c$ running from 30 K to 85 K in the various YBCO and Ca-substituted YBCO samples examined). The type of meso-scale activity which their observations would signal we earlier have witnessed directly in the electron microscope and recorded on video (unfortunately BetaMax) some 30 years ago for the case of the dc arrays in 2H-$TaSe_2$. 18 frames drawn from that video are included in ref. [60]. Typically these are spaced at 1 to 10 second intervals. The latter snapshots afford but a very limited glimpse of the activity witnessed on screen as the dc array fluctuates and pins at every possible level, from minute rapid adjustment to occasional avalanches. Flip-flopping of the dcs between more strongly pinned locations frequently was in evidence. The dcs are strongly susceptible to capture at their 'head' by some hidden pinning agent, and very often dc's would return to snag upon these same points during thermal cycling. Remember in the above *dark-field* images that *out-of-contrast* features (areal diamond strings bounded by dc's) run in the other two complementary hexagonal directions to define these singularities. It was found that as $\delta$ is altered slightly with temperature change new dc's would arise. The general perception was that the latter nucleated from defects and then would extend

'longitudinally', rather than for new dc's to 'compact in' from the edge of the sample 'transversely'.

The fact that all this activity in 2H-TaSe$_2$ manifests itself, as now is detected in YBCO, at remarkably low frequency levels is a marker of the extended nature of the fluctuators and of the significant size of the energy barriers being surmounted. By performing a Boltzmann analysis of the change in flip-flopping dwell-time ratios for an isolated fluctuator active still between 108 K and 100 K, Caplan *et al* have in [59] been able to evaluate a typical state free energy difference of about 37 meV (or 4½ *kT*). This energy stands a factor of 10 smaller than the typical activation energy associated with the general 1/*f* noise augmentation encountered up in the onset range from 260 K to 220 K. A Boltzmann activation energy of ≈ 0.4 eV was there determined, this coupled to an attempt frequency ~ $10^{10}$-$10^{11}$ Hz, suitably phononic in range and expressing the simple thermal kinetics of the extra 1/*f* noise generation. Despite the rapid increase in electrical 1/*f* noise met with near 250 K, the conductivity itself is less sharply altered, improving slightly as this activity is quelled and the striping becomes better organized. At lower temperatures the application of a magnetic field sees the noise level step up, due to further improvement to stripe organization, as more free energy becomes squeezed out of the material's magnetic subsystem under the spin-organizing field.

*4.5. The dynamics of stripe behaviour: the NMR/NQR studies of Suter et al. [61,62] on Y-124.*

The above noise studies of Caplan *et al* [59] are by no means the first time rapid onset of *charge* activity around 200 K has been identified. The NMR/NQR work of Suter *et al.* [61][62] back just prior to 2000 affords a case in point, of particular relevance now since it relates to YBa$_2$Cu$_4$O$_8$, the underdoped stoichiometric material playing such a key role in the 'quantum oscillation' saga. Suter *et al.*'s first paper disclosed, via the diverse measures of site Knight shift, relaxation rate and anisotropy, a unified spin-fluid active over all sites and exhibiting marked modification once below 200 K. The latter changes were demonstrated to be primarily not magnetic in nature, but of strongly quadrupolar (i.e. charge) origin. Because a change is evident too in field-free NQR data in addition to NMR, clearly this is not reliant upon the presence of a magnetic field (for the NMR work $H_{app}$ was 9 tesla). The NQR changes proved particularly pronounced both for the chain and the planar copper sites, and indeed for the O(4) oxygen between them, as would signal a J-T.-type lattice fluctuation involved with hole transfer from the chains to the planes. For the planar Cu(2) site both $K_c$ and $\nu_Q$ deviate sharply upwards below 200 K, whilst the relaxation rate at O(4) steps downward. The fluctuation time-scale was estimated as lying here in the range $10^{-9}$ to $10^{-5}$ sec. What fraction of this extra fluctuational activity is ascribable to magnetism and what to charge very cunningly was separated in the second of Suter *et al.*'s papers [62]. There through use of $^{17}$O-enriched (I=$^5/_2$) samples they examine the planar O$_{2,3}$ response at 9 T employing a novel double-irradiation technique. In this the central ($-^1/_2$,$^1/_2$) transition is saturated by the application of an appropriate additional stimulating r.f. field. A significant intensity *enhancement* is then uncovered for the ($-^3/_2$,$-^1/_2$) transition over the signal as recorded within the conventional spin-echo procedure. Normal spin-echo data are dominated by the $(-1/2, 1/2)$ magnetic fluctuations. (These were the ones pursued by Millis, Monien and Pines in their integrated treatment of the original NMR and neutron scattering data [63]). The Suter enhancement in the $(-3/2, -1/2)$ signal exhibits a strong rise from 200 K down to $T_c$, followed by a sharp collapse. The very marked $//c$ versus $\perp c$ anisotropy detected below 200 K in overall $O_{2,3}$ relaxation rate was adjudged to arise within the new quadrupolar relaxation channel. The latter's magnitude was appreciated as being far too high to issue from *individual* quasiparticle activity. Moreover it displays the wrong temperature dependence to be due to phonons. The above remarkably strong, collective, quadrupolar response from the planar oxygen sites speaks very strongly of stripe fluctuations, just as now do the new electrical noise data [59] – a topic that back in 2000 was not of foremost regard.

Why the above phenomena should appear so much stronger with $YBa_2Cu_4O_8$ than for $YBa_2Cu_3O_{7-\delta}$ becomes now an interesting matter (see [61] fig 12). Perhaps the stripes are better defined in Y-124 than in non-stoichiometric Y-123. As yet we have no neutron or synchrotron X-ray inelastic scattering work on Y-124 to affirm this. In the next section we shall examine what the role of the application of a magnetic field might be in this respect.

*4.6. dc jogging and anisotropies in diffraction: neutron scattering studies of Hinkov et al. [49].*

To finish this present section we insert next into the discussion very revealing inelastic neutron scattering results from underdoped YBCO-123, recently released by Hinkov and coworkers [49] and already alluded to above. These data relate to the discerned level of *a,b* basal-plane stripe anisotropy present in that system.

Extracting information from the neutron-probed susceptibility excitation spectrum $\chi''(Q,\omega)$ in HTSC material never is straightforward due to the multiplicity of processes which it enfolds. RVB-driven spin-coupling forces gapping to develop in AF spin-wave excitation extending right up towards 50 meV. Only beyond the latter energy does behaviour akin to that seen in the Mott-insulating parents become re-established. Below this energy residual spin excitation in fact remains in evidence but deriving from the more complex (circulatory) magnetic coupling associated with the striping arrays. The low energy fade out in stripe definition (as registered using neutrons) indicates this order too is spin-gapped excitation-wise. Many low-dimensional copper compounds experience comparable limitations upon their magnetic susceptibility imposed by their $S=\frac{1}{2}$ spin-count and by the high level here of *p-d* hybridization. These generic effects are now compounded by the superconductivity, pointedly as a consequence of this superconductivity having its own multi-component nature.

Beginning considerably above $T_c$, spectral weight becomes gathered into the 'magnetic resonance peak' centred upon $(\pi,\pi)$. Its energy of ~ 5–5½ $kT_c$ (41 meV for $YBCO_7$) associates it directly, within the strong-coupling scenario of the resonant crossover superconductivity, to an excitonic, pair-breaking, spin-flip in the condensate – global below $T_c$, but increasingly local above. The instigating local pairs lie themselves at a somewhat deeper binding energy (see fig. 2 in [10] and fig. 1 in [26a]): ~ –55 meV for the case of $YBCO_7$. At such an energy one encounters in addition the soft phonon-coupled mode sited in the saddle regions out toward ($\pi$,0), etc., and associated with local pair formation [64,10]. Those local pairs which remain outside the condensate manifest their own modal excitations. The first of these, as was discussed by Casas *et al.* [65], exhibits rather strong upward linear dispersion (i.e. to decreased binding), and it crosses '$E_F$' well prior to reaching the zone centre. This particular mode has, it is claimed (see fig.1 in [10] and figs.3,4 in [66]), been mistaken in many scanning tunnelling spectroscopy works [34,40] for the Bogoliubovons of a more standard superconductivity. A second collective mode, likewise of ($\pi$,0) starting point, but now showing much smaller dispersion, extends back into the long wavelength regions toward the zone centre: its origin has been discussed by Belkhir and Randeria [67]. This latter $A_{1g}$ symmetry mode is, as has been noted above, widely registered in electronic Raman [68], infrared [69] and ARPES [70] spectral work, it being strongly coupled with the $A_{1g}$ optical phonons [71]. Very recently this second, much more weakly dispersing, mode has been detected in addition in spin-polarized neutron scattering by Li and colleagues [72a] on $HgBa_2CuO_{4+\delta}$. The latter material, although very attractive to work with in many ways, has the disadvantage of jamming all these various modal features into a very narrow energy range. Li *et al* have examined both optimally and underdoped HBCO, and, as one might anticipate, find this mode to adjust to slightly higher binding energy with reduction in *p*. Its excitations then appropriately remain in evidence to somewhat higher temperature, in line with the customary behaviour of the pseudogap [66]. Alas this is not the final mode to demand untangling from the neutron scattering data because there still remains the 'downward' dispersing mode emerging from the resonance peak. The stability here of the *induced* d-wave pairs relative to $E_F$ is decreased on moving from the $\pi$,0 saddles and their associated pair creation 'hot spots'.

It is to scattering from this last mode to which now we direct attention for it rapidly overlays the scattering from the stripes. Indeed many have read the two as one unified process and have often lumped them together, along with the 'upward' dispersing spin waves (from approximately the same energy in YBCO), under the unenlightening label 'hourglass'. This unfortunate conglomeration of scattering does not help when seeking to find how the striping dissipates towards *higher* excitation energies, much as spin gapping does not help in viewing how striping might harden up towards *lower* ones. The new work from Hinkov *et al.* [49] using 94% detwinned $YBCO_{6.6}$ goes a considerable way toward clarifying these matters.

The chains in $YBCO_x$ extend in the *b* direction. For $YBCO_{6.6}$ diffraction spotting from the striping becomes first resolvable at energies below $E_{res}$ (here = 38 meV) and at temperatures once under about 230 K [see fig. 6a in 49]. This spotting is much the more apparent in the **a*** direction, corresponding to better defined diffraction 'objects' aligned in the perpendicular direction, i.e. parallel to **b**, the structural chain direction (see fig. 3a in [49] for *E*=23 meV at 3 K). With respect to neutrons those 'objects' are not the charge stripes themselves but the spin domains between them. In our figure 4 above we have shown that the counter-circulatory spin patterns in n.n. domains define a spin supercell of two interpenetrating fcc sublattices (as in NaCl) $8a_o \times 8a_o$ in size; i.e. the spin spotting is indeed rendered axial and with incommensuration vector δ ≈ 0.12**a₀**\*, etc.. We now can appreciate why the $a^*$-axis spotting in YBCO is so much more pronounced than the corresponding $b^*$-axis spotting (see [49], fig.3d). It is due not to an intrinsically 1D aspect to the source of diffraction, but to a superior level of organization of this $8a_o$ periodicity in supercell geometry when the cell edges and lattice chains run parallel. When mutually perpendicular the level of perfection of the supercell periodicity stands much poorer. A cell repeat that in the $b$ direction ought to be $9a_o$ readily fluctuates or 'jogs' to $8…10a_o$ (**figure 7a**). That means the degree of definition in the neutron scattering of *spotting* along $b^*$ is much the more diffuse (see [49] fig.17 bottom left). The bottom right panel within the same figure indicates the stripes still to be clearly in evidence at 70 K, i.e. somewhat above $T_c$ (61 K) for this underdoped sample. Indeed figure 6 in this very detailed and carefully worked paper tracks how the stripe diffraction intensity steadily fades away above $T_c$ only finally to disappear close to 200 K, perfectly in line with earlier discussion in this section. The paper permits the view that stripe incommensurability parameter $\delta$ can be held as being near invariant for all $T$ and $\omega$.

We will move forward now to look at the effect of high magnetic fields upon such stripe arrays.

## §5. The alteration wrought by application of a magnetic field on underdoped HTSC systems.

*5.1. Introduction to some problems within the present literature.*

It long has been known when one comes to measure the Hall coefficient for underdoped HTSC hole systems that $R_H(T)$ shows first indication of changing sign from positive to negative well in advance of the onset of superconductivity [73]. This is not the product simply of the magnetic field since it is a tendency that is in evidence too with the field-free measure of the Seebeck coefficient [74]. Furthermore it is not simply a move to the standard superconductive shorting out of all voltage signals, because the transfer becomes more marked whenever in a strong magnetic field able to suppress $T_c$ [75]. Pointedly $T_c$ likewise becomes significantly suppressed where striping is strongly frozen in, the classic case being LBCO in the vicinity of $p = ^1/_8$ although all HTSC systems indicate a comparable tendency [74c]. With striping involved in all these observations (as too in the many associated spin, charge and lattice diffraction effects [8,76,77]), it manifestly is not justified to disregard its presence and consequences when addressing the quantum oscillation data from underdoped HTSC material forthcoming at fields beyond 30 tesla [15-19]. This is the tack I followed in [27] and in this I am not alone. For example Norman and colleagues [20] turn to stripe formation to produce an electron pocket of about the perceived size. However, as stressed in [27], this element of reconstructed Fermi surface would issue from the axial saddles, precisely where the earlier-discussed incoherent behaviour is centred. This troubling result itself arrives from an adherence to Tranquada's long-held assertion that the charge- and spin-stripe ordering geometry is *1D* in form, with the stripes running basally in one or other principal *axial* direction [35a], an opinion hardened up through Kivelson and coworkers' analogy with nematic liquid crystals [35b].

As was pointed out above, our contrasting rotated 2D geometry has the benefit of setting the easy conduction mixed-valent domain boundaries in the nodal as against the antinodal directions. Such geometry sustains, furthermore, alternately sensed Cu(II) spin circulation within neighbouring domain interiors [33 fig. 2b] – see present fig.4.

One bar to hinder widespread acceptance of the latter view came with the observation made by Fauqué *et al.* [78] that despite underdoped YBCO-123 indeed supplying clear evidence, under probing by spin-polarized neutrons, of some unusual form of weak magnetic order, that order was in the first instance reported as bringing change only to the net intensity at the principal Bragg spots. These results clearly then cannot be ascribable to simple antiferromagnetism, whether of the Néel type or of the IC SDW variety. The adduced canting of the moments out of the basal plane and their small (site-averaged) magnitude (~0.1 $\mu_B$) indeed would suggest something more exotic. The initially perceived absence of change to the stripe satellite peaks prompted Varma to develop a homogeneous, on-coordination-unit interpretation of Kaminski *et al.*'s time-reversal symmetry-breaking magnetic circular dichroism data in terms of circulating micro-currents within each and every Cu-O coordination unit [53]. However the latest spin-polarized neutron work from Li, Balédent, Fauqué, Bourges *et al.* on Hg-1201/LSCO [72] would indicate that there do in fact arise changes in the satellite IC peak region, apparent once below ~ 250 K/120 K. Accordingly I feel happy to continue with the diagonal stripe domain model when assessing what is afoot in the QO experiments.

Before revisiting the QO results, it is important to incorporate next the results of two more recent papers coming from Tranquada and coworkers. The first of these relates to transport work on LBCO ($p = 1/8$) undertaken in magnetic fields of up to 9 tesla [79]. The second paper relates to the same system but now to structural and charge order work, conducted there under pressures of up to 3 GPa and employing synchrotron X-ray rather than neutron diffraction [80]. In §5.4 we then shall fold-in recent scanning tunnelling work from Beyer *et al.* [81] on optimally-doped YBCO-123 performed in magnetic fields of up to 6 tesla.

*5.2. The high-field results of Tranquada et al. on LBCO $p = 1/8$ [79a].*

As mentioned already above, the Seebeck coefficient in this material becomes negative in advance of a superconductivity that here in LBCO is only rather tenuously exhibited. Under a field of 9 tesla any trace of the latter quickly becomes suppressed to liquid helium temperatures to uncover a now much-extended range of negative $S$. What is emphasized with figure 1a in [79a] is that the transfer to negative values occurs sharply in this case at 54 K. This temperature is that at which the basic LBCO *crystal* structure converts from the LTO to the LTT phase under modified tilting of the Cu-O octahedra, converting from single-axis to double-axis geometry, and the space group changing from #64 (orthorhombic Cmca, or rather Abma) to #138 (tetragonal P4$_2$/ncm) [82]. The latter - note now *enlarged* - group contains both diagonal glide planes and screw diads. The new crystal structure would look to develop in response to the stabilizing diagonal striping – a point returned to in §§5.3 and 5.4.

Figures 1b and 1c in [79a] present the changes which arise in the basal and *c*-axis resistivities upon adoption of the LTT structuring. For both directions a small upward step is evident, as the Fermi surface becomes further trimmed under the supercell-induced gapping here experienced near $p = 1/8$. $\rho_{ab}$ at this temperature and composition is, note, ~ 130 $\mu\Omega$-cm, implying that, even as the fermions within the domains begin to localize, the carriers in the stripes themselves, both fermions and bosons, must remain still fairly mobile (see §3 above). Where this system really manifests marked conduction problems once below 54 K is in the *c*-axis direction. Conductivity in this direction would ordinarily be governed by the much superior band dispersion of the saddle states, but the latter states, of course, are precisely where at low temperatures the quasiparticles are being subjected to loss of coherence. Accordingly $\rho_c$ climbs very steeply upon cooling, and, as is to be seen from fig. 1c of [79a], comes by 10 K in a field of 9 T to reach ~2 $\Omega$-cm, a most un-band-like value. Correspondingly for YBCO$_{6.67}$ ($p\approx 1/8$) no *c*-axis far-IR plasma response is in evidence optically at 8 K, whether at 9 T or 0 T [79c], and likewise with underdoped LSCO and BSCCO [83].

*5.3. High-pressure results of Tranquada and colleagues on LBCO $p \approx 1/8$ [80].*

Frequently one hears voiced the criticism that stripes only materialize in LBCO ($p\approx 1/8$) with the crystal structure shift from the LTO to the LTT phase, as against the option that the LTT phase marks compliance of the lattice to the growth in charge ordering, 2D in form. At room pressure static charge order and the structural change do indeed arrive at the same onset temperature of 54 K. Under applied pressure, however, it now is found possible very clearly to demonstrate that the charge order, once beyond a critical pressure of $P_c = 1.77$ GPa, actually is able to persist through into the HTT structure, both LTT and LTO structuring being eliminated beyond this pressure for all *T*. In this illuminating high pressure work the *charge* stripe order has been sensed directly by turning to synchrotron X-ray diffraction as opposed to neutron diffraction, a course of action that earlier was followed by Niemöller *et al* [84] in their examination of *x*=0.15 (La,Nd,Sr)$_2$CuO$_4$ (see discussion in §3.4 of [32]).

In the charge-ordered condition within the HTT crystal phase being now accessed in LBCO $p=1/8$ above $P_c$, it is revealed in [80] that, although the diffraction does become somewhat more diffuse, a correlation length ~80 Å nonetheless remains in evidence even at 2.7 GPa. The $T_c(P)$ and $T_{CO}(P)$ plots look set to intersect at 15 GPa, by which pressure (as can be ascertained directly) $T_c$ will have risen to 18 K (see inset to fig. 1 in [80]). The latter, notwithstanding, remains little more than half the $T_c$ value exhibited by LSCO $p=1/8$, for which stripe formation is less strongly frozen in.

*5.4. Beyer et al.'s STS evidence of magnetic-field-enhanced stripe order in YBa$_2$Cu$_3$O$_7$ [81].*

The natural effect of a magnetic field upon the HTSC systems, as has been demonstrated using neutron scattering by both Lake *et al.* [77] and by Gilardi *et al.* [85], is to promote decoupling of RVB spin-pairing and to bring transfer toward an ever more local spin state. The change entails sharper differentiation between those sites sitting within the interior of a stripe domain and those on the stripes themselves. The former become more consolidated as $^9Cu_{II}^0$, with the $^8Cu_{III}^0$ holes – plus their negative-$U$ bosonic double-loadings $^{10}Cu_{III}^{2-}$ – segregated to the mixed-valent domain boundaries. The condition is illuminated using the neutron diffraction probe, which under increasing $H$ reveals a fall in the spin gap magnitude and a build-up in antiferromagnetic coupling and (spiral) spin-ordering, conveyer of the earlier noted rise in $\rho$. Additionally it precipitates rather rapid depression of the superconductivity over a sizeable composition range around $p = {}^1/_8$ as commensuration effects mount.

As $p$ is increased beyond 0.125, the (average) stripe periodicity has to adjust from $8a_o$ to $7a_o$, but the latter commensurate geometry is not actually gained until the hole content $p$ exceeds 0.156 (see fig.4 in [32]). Gilardi *et al.* examined $p$=0.17 LSCO and there a $7a_o$ superlatticing clearly had become established [85]. The new paper from Beyer *et al.* [81] deals now with fully oxidized YBCO$_7$ ($T_c$=93 K), for which $p \approx 0.165$. This scanning tunnelling spectroscopy (STS) work encourages one likewise to proceed with the $7a_o$ superlattice arrangement illustrated earlier in figure 5, wherein ideally $p = {}^8/_{7^2}$ or 0.1632.

Let us examine figure by figure what Beyer *et al* actually record, free initially of interpretation. Their fig 1a is a standard zero-field plot of tunnelling conductance versus voltage bias at three selected temperatures, 6 K, 77 K and 102 K. The latter trace from ~10 K above $T_c$ is here virtually featureless. The other two traces each display a pair of gaps. The primary gap of magnitude $2\Delta = 40$ meV, being of the superconducting gap size, accordingly carries the gap parameter label $\Delta_{sc}$. At almost twice this energy there exists a shoulder, given here parameter label $\Delta_{eff}$. Of these two zero-field features, $\Delta_{sc}$ is much the more spatially homogeneous, exhibiting a spread in values half that found for $\Delta_{eff}$ (fig.1bc and 1ef). The two features, nevertheless, bear such intimate interrelation that one may presume to decompose $\Delta_{eff}$ through $\Delta_{eff}^2 = \Delta_{sc}^2 + V_{CO}^2$, with $V_{CO}$ (=32 meV) being taken to refer to a coexisting 'density wave' state.

Under a magnetic field of 2 tesla spectral changes arise, the form of which very much depends upon whether one is scanning within a flux vortex or between vortices (fig.1d). The $\Delta_{sc}$ gap alone is witnessed to arise within the latter regions, whilst the larger composite $\Delta_{eff}$ gapping feature is measured inside the vortices, this distinction being rather sharply demarcated (see figures 2c,d). Figures 2a ($H$=2 T) and 2b ($H$=4.5 T) disclose that the spatial array of flux vortices is itself rather disorganized here at 6 K under applied fields of these magnitudes, the array already not holding to the classic regular hexagonal form. The vortex haloes clearly are larger (~50 Å at 6 T) than is expected from the assessed superconducting coherence length $\xi_{sc}$. Furthermore the individual conductance spectra acquired from a wide selection of spatial pixels never supply any indication of a zero-bias peak of the form seen with conventional superconductors. Figure 4 emphasizes further the non-conventional nature of the STS data. There, while the individual gap energies $\Delta_{sc}$ and $V_{CO}$ themselves alter rather little with field strength (they run parallel), their relative spectral weights quickly crossover ($I_{shoulder}$ increasing as $I_{sc}$ falls). This is not just due to spectral broadening, because the two features each broaden in identical manner with $H$ (fig.4c). A considerable amount of the above is not entirely unexpected and has in part been reported previously in the early STM work of Maggio-Aprile *et al.* [86], etc. (see also Fischer *et al.* [87]). What is really new and important now in Beyer *et al.*'s paper all has been crammed into the diverse parts of figure 3, the unravelling of which will make evident why the above energy label $V$ carries the suffix CO for 'charge order'.

Figure 3 of [81] reports a Fourier transform study in basal reciprocal space obtained from *spatially-integrated* conductance data and it entails *difference mapping* between the field-applied and field-zero conditions. The primary tunnelling data are gathered from an area roughly 1000 Å square at 6 K under fields of 0 and 5 tesla (at which field the vortex spacing has just dropped below 200 Å). The FT-derived **q**-spot patterns of figures 3a, b and c refer respectively to tunnelling energies $\omega$ of (*i*) −10 meV, (*ii*) −20 meV, and (*iii*) the *energy-integrated* tunnelling signal from −1 to −30 meV. Recall here that $\Delta_{sc}$ = −20 meV. The upper parts of figures 3d and 3e relate to *anti-nodal* (axial) behaviour in the various plots given therein, whilst their lower parts relate to the corresponding *nodal* (diagonal) behaviour. What actually are presented in figure 3d of [81] are intensity slices from three FT difference maps such as those of figures 3a-c, traversing first axially from $\Gamma$ to $(2\pi,0)$ and then secondly diagonally from $\Gamma$ towards $(\pi,\pi)$. Next in figure 3e the various **q**-peaks featuring in these traces are tracked individually as functions of the tunnelling bias energy $\omega$ down to about −40 meV. These plots disclose both dispersive and non-dispersive components. Finally figure 3f selects the *non*-dispersive component signals and examines their symmetry characteristics with regard to bias voltage *sign* - whether they are symmetric or anti-symmetric. Remember all the above relates to the FT *difference* maps for the changes incurred between 0 T and 5 T.

From this plethora of observations let us begin with two features which we shall not dwell on subsequently. Firstly there is the strongly energy-dispersive response of the nodal feature, vector $\mathbf{q}_7$. That was the feature originally concentrated upon by Hoffmann, McElroy and colleagues in their STS work of [88] and subsequently generalized in [89,90]. The discussion by those particular workers was in terms there of Bogoliubovon modes and scattering interference between symmetry-equivalent positions spaced apart on the Fermi surface by such wavevectors. In contrast, the current author elected to attribute these $\omega(\mathbf{q}_i)$ dispersive modes to local-pair bosons excited outside the condensate (see [10],[28b] and [91]). Within the above FT maps there next occurs the strong feature present at $(\pi, \pi)$, etc., and which is clearly associated with the spin-pairing singlet-triplet spin-flip. The latter is, one recalls, much enhanced under the applied 5 tesla field. For our present purposes where real interest lies now is with the residue of *non*-dispersive, periodic features found to dominate the *intra*-vortex signal, and accordingly too strongly *field-enhanced*.

What then are the wavevectors associated with the three, *non*-dispersive, field-promoted, charge/spin/structural modulations sensed here within the vortices and manifesting such direct response to the site tunnelling currents? Using the pseudo-tetragonal basal lattice constant, *a*, for convenience, the following are the wavevectors as given in [81]:

axially -     $\mathbf{Q}_\alpha = (0, 0.28 \pm 0.02) \pi/a$ ,   and   $\mathbf{Q}_\beta = (0, 0.56 \pm 0.06) \pi/a$ ,

diagonally -    $\mathbf{Q}_\gamma = (0.15 \pm 0.01, 0.15 \pm 0.01) \pi/a$ ,     (together with symmetry equivalents).

Rewriting these more succinctly and informatively we have

$\mathbf{Q}_\alpha = (0, 0.14) 2\pi/a$ ,      $\mathbf{Q}_\beta = (0, 0.28) 2\pi/a$ ,      $\mathbf{Q}_\gamma = (0.15, 0.15)½ \cdot 2\pi/a$ .

Immediately within the accuracy of measurement one recognizes that this is

$\mathbf{Q}_\alpha = (0, ^1/_7) 2\pi/a$     and    $\mathbf{Q}_\beta = (0, ^2/_7) 2\pi/a$       (axially directed),

with    $\mathbf{Q}_\gamma = (^1/_7, ^1/_7) ^1/_2 \cdot 2\pi/a$                (diagonally directed),

making    $|\mathbf{Q}_\gamma| = ^1/_7 \cdot (^1/_{\sqrt{2}} \cdot 2\pi/a)$        (see FT cut in lower fig.3d of [81]).

Accordingly    $\lambda_\alpha = 7a$ ,  $\lambda_\beta = 3½a$     , both axially directed,

and    $\lambda_\gamma = 7 \cdot \sqrt{2} a$              , diagonally directed.

There can be little doubt then here that we are dealing in optimally-doped YBCO-123 at 4 K and 5 T with a 7*a* x 7*a* superlattice. The task is to identify the precise nature of the current modulations $\alpha$, $\beta$, and $\gamma$.

Bearing on this we have the above mentioned bias-voltage $\pm\omega$ symmetry status to consider beyond simply the wavevectors themselves. Beyer and coworkers [81] find the FT of the difference signal at $\mathbf{Q}_\gamma$ to exhibit a large antisymmetric component to the tunnelling intensity with repect to bias voltage sign. By contrast $\mathbf{Q}_\beta$ is dominantly symmetric wrt $\pm\omega$, whilst $\mathbf{Q}_\alpha$ is symmetry neutral. Thus we see that $\mathbf{Q}_\alpha$ (in part) and $\mathbf{Q}_\gamma$ relate to the true spin array period within the field-enhanced diagonally-striped structure, whilst $\mathbf{Q}_\beta$ is identified as the charge wavevector for the patterning, portrayed above in figure 5.

**§6. The oscillations detected in underdoped HTSC samples are dictated by real space and not reciprocal space physics.**

*6.1 Recap of where we have reached, including a first look at the QO results.*

§3 showed that recent transport data on underdoped HTSC materials are better treated by a two-component scenario that is based on hole quasiparticles (from the nodal regions) and local pair bosons (created from the electrons in the saddle regions of the basic FS), rather than by two oppositely signed sets of fermionic Landau quasiparticles drawn from a reconstituted Fermi surface with attribution to SDWs, DDW's, CDW's/stripes, or whatever.

§4 affirmed the existence of fluctuating stripes in all these systems. The drive to stripe formation is valence segregation, not Fermiology, and the geometry of the array being

settled into is 2D in form, not 1D. The resulting Skyrmionic-like (essentially antiferromagnetic) spin array sports a very small overall moment plus a spin canting under a lattice coupling which emanates from Jahn-Teller induced strain at the stripe boundaries. The stripes settle in slowly below ~ 200 K and are a manifestation of the pseudogapping. The latter, however, is rather more fundamentally based, resulting from on the one hand proximity to the Mott transition and on the other from dynamic local pair generation in the prevailing B-F crossover circumstance. Only at lower temperatures ~130 K does a substantial, quasi-static, local pair population establish itself, capable of generating the observed Nernst effect [92].

§5 showed that stripe formation becomes more pronounced in a strong magnetic field. The fermions become more local in character, and that finds expression both in transport and magnetic behaviour reminiscent of the changes wrought by Zn substitution [93]. At fields above 30 T, the stripe domain interiors begin transferring into the normal state, with the mobile pairs more and more confined to the stripes and the residual coherent quasiparticles to the F.S. nodes. Incoherent fermions join the uncondensed local pair bosons as *negative* quasi-classical particles.

Upon this localizing milieu now have been unleashed the 'quantum oscillation' techniques, the outcome regarding which I already have forcibly expressed my misgivings in [27]. There I indicated how the observed $1/B_m$ oscillatory data could arrive via a very different route to that advocated in [14-19] and subsequently in [24,25,94-96]. The proposed real space route invokes the stripe array affording a ratchet release to the passage of flux vortices across the stripes themselves as the field strength is changed. Besides the $1/B_m$ form upheld in this modelling of the oscillations (running numbers m) and the general magnitude of the evaluated peak series 'head fields' $B_{m=1}(p')$ being close to those signalled experimentally, a change in $B_1(p)$ linear in sample doping $p$ also was supported. A strict linearity between $B_1(p)$ and $p$ would, if experimentally affirmed, speak strongly against any fermiologically based scenario since the detail of the reconstructed Fermi surfaces at different $p$, and in particular from different systems like 123 and 124, never is going to be so simply and uniformly constrained. Remember in this regard the as-reported F.S. pocket cross-sections all emerge consistently as only ~2% of the parent Brillouin zone area.

In the postscript to [27] I pointed to Audouard *et al* [19b] having just made report for YBCO$_{6.5}$ of a basic $B_1(p)$ peak being flanked by two lesser peaks, and that these three would support a linearity $B_1(p) \propto p$ if the fundamental peak for $p = 1/10$ were to be regarded as straddled by secondary signals from $p = 1/11$ and $p = 1/9$. An analogous triplet set of $B_1$ values of 460, 532 and 602 T more recently have been reported by Sebastian *et al* [94] for a higher $T_c$ 123-YBCO$_x$ and these likewise would support a simple linear relation, appertaining now to $p = 1/10, 1/9, 1/8$ were the quoted fields to be taken as 478, 531 and 596 T – these latter are deemed to lie within reasonable experimental bounds. The above flanking features might be comprehended as due to doping inhomogeneity within the sample, but more likely are due to jogging within the striping array, as illustrated earlier in fig.7a. Such jogged segments occur far more readily in a direction perpendicular to the chains in YBCO$_x$ than parallel to them, as signalled in §4.6 regarding the recent neutron work from Hinkov *et al.* [49]. As is illustrated in fig.7b the resulting adjustments to the domain geometry will yield domain areas/hole contents interrelated in the above fashion. It is felt such sourcing of these widely reported triplet features is far less forced than it is being invited to rely upon *c*-axis variance in the cross-sections of the two proposed Fermi cylinders, especially when it is required that two of the four extremal areas be identical. Furthermore a couple of the adduced effective masses also emerge from the out-workings of the Lifshitz-Kosevich treatment as identical, viz. ≈ 1.6 $m_e$ , – a value well below that previously found even in highly *over*doped samples [14]!

Despite such pitfalls being sidestepped by the real-space ratchet scenario ventured in [27], something now is appreciated as demanding amendment within that exposition. This relates (in part) to the *p*-values employed in its development - and in particular for YBCO-124. Although the three $B_1(p)$ values figuring in table 1 of [27] express a linear progression with *p* – reflecting in this the way they were evaluated – they do not afford the desired extrapolation through the origin; the (implied) gradient is too steep. To begin with it is evident *p'* for YBCO-124 requires pushing up to around $1/8$ from my originally inserted value of $1/9$.

It is most unfortunate in this work that we have to deal with the two YBCO systems, for which there are the complications interposed by the chains in arriving at the *p* values specific to the planes. Reliance upon the $T_c(p)$ plot underscores this problem – the marked dipping evident here, as in most of the HTSC systems, occurs over precisely the underdoped *p* range currently in question. It will be of benefit now to examine some of this 'unwanted' detail more closely.

*6.2. $T_c$ dips, stoichiometries, precise p-values, and stripe periodicities.*

Hücker *et al.* have presented recently in [55] a detailed study of the stripe order in (La$_{2-x}$Ba$_x$)CuO$_4$ across the range 0.095<*x*<0.155 in which striping is most in evidence and where the dip of $T_c(p)$ away from simple parabolic variance arises. Hücker *et al.*'s X-ray and neutron as well as static magnetization measurements, all undertaken on high quality single crystal samples, permit the developments in charge and spin ordering now to be better assessed. For all compositions within the dip range the onset of frozen charge order occurs at slightly higher temperatures than does the spin order. Recall that in LBCO this *frozen* order is enmeshed with the basic crystal structure transformations LTO/LTLO → LTT [55, fig. 10]. What at this stage I wish to emphasize is presented in **figure 8**, an amplified version of fig.15 in [55] containing their empirical findings as to how $x_{Ba}$ is related to $\delta(x)$, the incommensurability *k*-space parameter (as a fraction of $2\pi/a_o$). Because of the simple way in which the hole content is generated in this system, there is little problem here in equating $x_{Ba}$ with *p* (provided one ensures that the oxygen content is kept strictly at -O$_4$). Figure 8 strikingly discloses that, in spite of received wisdom, there is no lock-in here at $p = 1/8$ to $\delta = 1/8$. Indeed a $\delta$ value of $1/8$ is not in fact reached much prior to $p(T_c^{max})$. The data in fig.8 would for $p = 1/8$, by contrast, support a $\delta$ value of 0.118. That number amounts to ($1/8+1/9$)/2, one earlier in evidence with LSCO and HBCO as being the *p* value at which $T_c(p)$ is most depressed — not 0.125. One observes from fig.8 that the actual doping level for which we do attain the simple relation $\delta(p) = p$ is (from this data at least) $p = 1/9$. Thus our earlier figure 4 was not describing reality here, whilst figure 6 would seem to be. Such a result could suggest for the 'QO'-directed scenario proposed one ought to employ as unit area, $D(p)^2$, not the single domain size but perhaps twice this (≡ *axial* supercell area), if not 4 domains, bearing in mind the true charge arrangement and spin chirality repeat pattern. The monotonic form to $\delta(p)$ in evidence in fig.8, however, would imply that the more important parameter here is not so much the supercell area $D(p')^2$ as the supercell hole count $N(p')$. One observes from figure 8 that just as point ($p = 1/8$, $\delta = 1/8$) did not sit on the empirical curve, neither too does ($p = 1/10$, $\delta = 1/10$) — again a popular misconception. Where in fact $\delta$ does equal $1/10$ is for $p = 9/10^2$, at which hole loading, as may be seen from **figure 9**, a regular site-alternation within the stripes is procured for accommodation of these 9 holes per cell.

How much more intricate now might one expect the situation to be in chain-bearing YBCO-123? The indication from **figure 10** is considerably so. That figure plots out YBCO data abstracted from table 1 in a recent paper released by LeBoeuf *et al* [97] on high-field Hall measurements. It manifests the anticipated break in behaviour relating planar hole-fraction, *p*, to the net oxygen stoichiometry $-O_x$, – defective within the chain region of the structure. *p* has been gauged here via the customary practice of measuring $T_c(p)$ together with its assayed departure from the idealized parabolic form set by the wings of the curve, a procedure first formulated by Presland and Tallon [98]. **Figure 11** displays the empirical $T_c$ vs $-O_x$ information collected by LeBoeuf *et al* in [97], plotted up now on the same scale as figure 10. Figures 10 and 11 are each so significantly different in gradient at oxygen contents below 6.50 and above 6.66 as to warrant the view that phase immiscibility arises here somewhere between $-O_{6.56}$ and $-O_{6.66}$. From figures 10 and 11 the two-phase region is adjudged to be centred precisely on where $p = 0.118$, the mean of $1/9$ and $1/8$. This state of affairs would account for why so many researchers specify their $YBCO_x$ sample stoichiometry as being $-O_{6.56}$. We have come to exactly the same point then as with the 214 family. No monotonic quantity, let alone linear quantity, can run through this range of underdoping other than *p* itself. Although $D(p)^2$ and $N(p)$ of course are intimately related, $N(p)$ is accordingly set to be the primary quantity as regards the physics. This conclusion will be something to watch out for now as we turn finally to consider the 'quantum oscillation' results.

*6.3. Remarkable consequences regarding the conditions behind the 'quantum oscillation' results.*

It is necessary first to appreciate just what was being proposed in my original treatment of this matter in [27]. Below I have set this out carefully in **tabular form** so as to be able to refer precisely to where it now stands in need of revision.

At this point I introduce into development of the argument the collection of $B_1(p)$ data contained in a review of QO work just published by Vignolle *et al* [99]. There a set of $B_1(p)$ points is presented, seemingly linear in $p$, linked by the dotted line and now reproduced here in **figure 12**. This plot incorporates a point for Y-124. The latter, however, when as in [99] ascribed to $p = 0.14$, drags this line away significantly from extrapolation through the graph's origin. As has been expressed above and earlier in [27] I believe the above $p$-value for Y-124 (additionally employed by the authors of [16]) is, however, appreciably too large. While the $p$ value of 0.11 I was drawn into using in [27] is definitely too low, I am persuaded still to drop the value substantially to 0.128. That now will permit the line through the $B_1(p)$ data points to extrapolate to the origin as desired. A quite underdoped $p$-value like 0.128 stands, as was emphasized in [27], in better keeping with the very large pressure coefficient $dT_c/dP$ exhibited by Y-124, for which $T_c$ rises from 80 K to 110 K under 30 kB [100]. In regard to the further point positioned in [99] at $p = 0.12$, a value of 0.116 would place that data point within the awkward mixed-phase regime (and close to the revised line!). The gradient for the new origin-incorporating $B_1(p)$ line appearing in figure 12 takes on now the value of 5170 tesla per planar Cu. Before showing that this is a particularly significant value, we should return to **table 1** and examine again the course of action I pursued in [27].

In that table equation (1) serves to define $N(p)$ and $D(p)$, while equation (3) defines $M(p)$ and equation (5) defines $\mu(p)$. Equation (7) introduces the extrapolated 'head fields' with m=1 which top the experimentally detected sequences of fields $B_m(p)$, $B_1(p)$ being here deemed directly proportional to $p$. That was presumed in [27] and one judges it affirmed now in the collected data. In [27] I crucially also presumed $\mu_1(p)$, the number of fluxons per 'superunit' under fields $B_1(p)$, to be unity. With this latter assumption all fields $B_m(p)$ would stand, as evaluated, about a factor of four below those observed or extrapolated from the 'quantum oscillation' experiments [15-19] - *upon* (i) the applicable area $D(p)^2$ being taken as the entire supercell area and (ii) the standard flux quantum, $\Phi_o$, being adopted. In [27] in order to 'recover' this factor of 4, I accordingly transferred from the full supercell area to the stripe domain area (a first factor of 2) and then employed as flux quantum $2\Phi_o$ rather than $\Phi_o$. The latter change secured the second factor of 2, it then being necessary for one to have to go to twice the applied field to accumulate the requisite number of vortices appropriate to operate the flux ratchet process and match the observed data. When proceeding in this way, as is to be seen from table 1 in [27], it was found necessary additionally to drop the $p$ value taken for YBCO-124 down to 0.111. This latter 'adjustment', as signalled above, I now am persuaded is somewhat excessive. A value of 0.128 would appear much more appropriate because as noted, the 124 point then sits on the heavy straight line in figure 12 through Vignolle *et al.*'s collected data [99] holding extrapolation through to the origin.

Let us back-track at this point concerning the assumption that as regards a *domain* $\mu_1(p') = 1$, and revert to the relevant area $D(p)^2$ as being the *entire* area of the stripe supercell, and furthermore retain standard flux quantum, $\Phi_o$. In equation (8) of table 1 we now pursue the alternative prescription to making $\mu_1(p')$ unity of instead accepting that $\mu_1(p') =$ ($c_1/\Phi_o$).$N(p')$. The number of fluxons per cell $\mu_1(p')$ then is fixed not so much by the cell area $D(p')^2$ as by its actual hole content $N(p')$. (N.B. $N(p) = p.D(p)^2$, with $D(p)^2$ here in nm$^2$). With $c_1$ experimentally set as 5170 tesla per Cu ($p=1$) and with $\Phi_o$ = 2067 tesla nm$^2$, this leads to the striking outcome $\mu_1(p')/N(p')$ = 2.50. One now is encouraged here to replace 2.50 by integer fraction $\kappa = {}^5\!/_2$ (whereupon $c_1$ becomes 5167) and to assert that, very remarkably, we look to be making contact with the physics of the fractional quantum Hall effect (FQHE). Within that topic $\kappa = {}^5\!/_2$ flux tubes per relevant charge constitutes a fairly stable circumstance appropriate to Abelian 'filling factor' $\nu = 1/\kappa = {}^2\!/_5$. We attempt below in §7 to understand how such a situation could come about within the present context.

*6.4. The matching of the original data to the new appreciation.*

Before closing this section, however, it is vital now to affirm how well this new assessment of the action indeed copes with accommodating the actual 'QO' data. **Figure 13** records how $B_m(p)$ steps down through the integer m series at each $p$ under the above ${}^5\!/_2$ prescription for the 'head fields' $B_1(p)$. (Just how such a stepped $^1\!/_m$ sequence continues to materialize will be dealt with shortly.) The way one actually is able more readily to relate to the experimental data is in fact to plot $1/B_m(p)$ against $p$ (or better $N(p)/D(p)^2$), as made clear by **figure 14**. The universal curves there become equally spaced at any given $p$, and thus by measuring the average spacing in the experimental traces between successive oscillations it immediately becomes possible closely to ascertain a sample's $p$ value. If we take the example of Jaudet *et al.*'s YBCO-123 $1/B_m(p)$ data [18] as basis of the present **figure 15**, we see the trace from its periodic spacing of 0.00185 T$^{-1}$ achieves accommodation in figure 14 at $p$ = 0.1045. Upon reference to figure 10 the latter $p$ value is, one finds, appropriate to an oxygen content of 6.53 rather than to the 6.51 cited by Jaudet *et al.*. From figure 11 their stated $T_c$ value of 57 K similarly would lie 4 K lower than the value expected of the higher -O$_x$ value. The 'head field' of 540 T (=10/0.01850) is however identical to what was cited in [18] from the Fourier analysis of their oscillatory trace. It would be good now to have reliable results for even lower $p$, and Sebastian *et al.* in [101] recently have reported work in this direction. However, when we come to examine the extensive experimental trace in the upper inset to their figure 2B, claimed as being for -O$_{6.50}$, that data with its spacing of 0.00210 T$^{-1}$ would appear upon proceeding in the above manner to relate $p$ = 0.0975, appropriate from our figure 10 to -O$_{6.515}$. (The relevant numbers are $B_{10}$=50.50 T ($B_{10}^{-1}$=0.01980 T$^{-1}$); $B_9$=56.12 T ($B_9^{-1}$=0.01782 T$^{-1}$); $B_8$=63.13 T ($B_8^{-1}$=0.01584 T$^{-1}$); $B_7$=72.15 T ($B_7^{-1}$=0.01386 T$^{-1}$); $B_6$=84.17 T ($B_6^{-1}$=0.01188 T$^{-1}$)). The above combination of $B_1(p)$=505 T, $p$=0.0975 and $x$=6.515 is not indeed as represented in [101, fig.7A], but it does accord precisely with the lowermost point in the current figure 12 containing LeBoeuf *et al.*'s values [97].

Turning to the more contentious case of Y-124, **figure 16** presents the 'QO' data of Yelland *et al.* [16b] as marked up now with the appropriate $1/B_m$ values supplied by our figure 14 for the experimentally observed oscillation spacing 0.001515 T$^{-1}$. The latter directly specifies $p$ for YBa$_2$Cu$_4$O$_8$ as being 0.1277. The 'head field' is identified as being 10/0.01515 or 660 T, and once more this matches the value in [16b] of 660±15 T extracted by Fourier analysis of their limited data set. The above emphasizes then how all the above numbers become dictated in a very direct manner without manifesting any of the variability inevitable under a nesting scenario.

The Y-124 data in [16b] from Yelland *et al.* and in [16a] from Bangura *et al.* were in figure A.2 of [27] presented alongside each other, and that figure draws attention now to an important feature of these various oscillatory traces not emphasized so far; namely their relative phasing at head field $B_1(p')$. The relative trace phasings alter with the actual techniques and properties used to supply the information – $\Delta\rho/\rho$ with the de Haas-Shubnikov measurements of Bangura *et al.*, or the frequency change in the inductance measurements of Yelland *et al.*, or the magnetization in de Haas-van Alphen measurements, etc, etc.. We note in Bangura's 124 work that the above identified fields reveal their $\Delta\rho/\rho$ traces to be of simple sine-wave phasing ($\psi$ =0) (upon going from low field to high field). Thus when reading their $B_m$ trace *from left to right* they specify '– to +' crossover points, but when so reading a $1/B_m$ trace they would specify '+ to –' crossover points. By contrast the inductance technique of Yelland *et al.* sees these same $B_m(p')$ in their $\underline{1/B}$ trace fall at '+ to –' crossover points in their $\Delta f$ signal; i.e. the data display phasing $\psi = \pi$ relative to the $\psi = 0$ sinusoidal phasing of the $\Delta\rho/\rho$ data. Note Bangura *et al.*'s $\underline{B_m}$ trace appearing in [16a,fig.2] runs from m=13 '+ to –' crossover field at its left-hand to m=11 at its right-hand, whilst Yelland's $\underline{1/B_m}$ trace in [16b, fig.2] runs from '+ to –' crossover field m=8 at its left-hand end through to m=12 at its right. The relative phasing shifts between these various traces accordingly are seen to carry interesting information regarding the physics of signal generation, a matter we return to in §7.

*6.5. Developments in QO work from Ramshaw et al. [102] and Riggs et al. [103].*

It is time now to look at a couple of more recent 'QO' papers that bring something new and significant to the problem, released very recently by Ramshaw *et al.* [102] and by Riggs *et al.* [103]. The former is of the de Haas-Shubnikov ($\Delta\rho/\rho$) type and upon examining the bottom trace in figure 3 of [102] we see a $1/B$ trace with spacing ~0.00186 T$^{-1}$, with $B_{10}$ specified by the first '– to +' crossover point from the left. That would locate $B_1$ in the vicinity of 535 T, making $p$ near 0.103 and $x$ near 6.53, very close to the situation with Jaudet *et al.*'s results of [18] discussed above. This is not too surprising as the new results come from the same group. Ramshaw *et al.* once again point out that their trace manifests, through its beats, the added presence of two weaker components having $B_1$ near 475 and 600 T, the triplet situation we have drawn attention to already. Ramshaw *et al.* have collected a much wider bank of data than before, including using a range of sample tilt angles of up to 55°. With this information the authors attempt a novel global fit to their entire data set, adhering in this to the regular Lifschitz-Kosevich formulation. One outcome of the analysis is that the dominant $B_1$ component emerges with a slightly downshifted value of 526 T, the subdominant component then being cited as $B_1$ = 478T. Their tilt experiments were triggered by the notion that the signal splitting arises from *c*-axis warping of the F.S., as was commented upon earlier. However I would once more point out that, as with Sebastian's work, the triplet of $B_1$ values 473.5:526:584.5 takes the integer ratio 9:10:11. Ramshaw *et al.* proceed next to consider the signal nodes manifest under sample tilting, all as based upon a fermiological interpretation, with the changes witnessed here ascribed to field-induced spin-splitting for the two components dealt with. The match to the data is not good at high tilts. It leaves one feeling this is not the direction to pursue. The de Haas-Shubnikov measurements actually were conducted in this experiment with both $\rho$ and *H* parallel to *c*. $\rho_c$ is barely coherent in underdoped samples even in zero field (0.5 $\Omega$-cm at 40 K), and as we have observed a strong field strengthens striping and localization greatly. Wen *et al.*'s recent study on $p = 0.092$ LBCO [104] serves to emphasize what a lethal combination the above is for underdoped material with respect both to normal state *c*-axis quasiparticle behaviour and to any interplanar Josephson tunnelling of pairs. Perhaps the most important conclusion that Ramshaw *et al.* [102] procure from their study of the contributory Zeeman splitting factor to the overall oscillation amplitude analysis is that the relevant spin *g*-factor is close to 2. Why this is significant is because with *H*//*c* as here, it cannot relate to the majority of sites in the interior of the domains if the spins are locked by the 'SDW' into the planes, as indicated in our figures 4-6 and 9. For these *g* would be close to zero. This throws responsibility then for the *g*=2 'free-spin' response being registered here upon the unconstrained spins within the stripes themselves.

The recent paper from Riggs *et al.* [103] (conducted upon YBCO-123 of very comparable oxygen content to Ramshaw *et al.*'s) proceeds now in making very apparent that the above standard view is indeed unsupportable. It introduces something really new by looking not at transport or magnetically related oscillations, but by going to the heart of the free energy problem to monitor field-induced oscillations in the specific heat. Once again this new work uncovers two components to the overall signal, but of a very different and highly illuminating form to be examined now in §7.

### §7. Regarding attainment of the FQHE state with filling factor $\nu = {}^2/_5$.

*7.1. The compound behaviour of the electronic specific heat in the study of Riggs et al. [103].*

The spacing in Riggs *et al.*'s [103] oscillatory $\Delta C_p^{el}(H)$ trace is at 0.00188 T$^{-1}$ very close to that above in Ramshaw *et al.*'s [102] $\Delta\rho/\rho$ trace. In figure 2B of Riggs' paper the top (1K) *B* trace runs from m=17 at the left to m=12 at the right, and, note, the relevant fields now specify $C_p$ *minima*:

| m | 17 | 16 | 15 | 14 | 13 | 12 | 11 | & 1 |
|---|---|---|---|---|---|---|---|---|
| $B_m$ | 31.19 | 33.13 | 35.34 | 37.85 | 40.75 | 44.13 | 48.12 | 531.9 T |
| $1/B_m$ | 0.03206 | 0.03018 | 0.02830 | 0.02642 | 0.02454 | 0.02266 | 0.02078 | 0.001880 T$^{-1}$. |

With this 0.00188 T$^{-1}$ spacing, the associated values of *p*=0.103 and *x*=6.53 again stand as slightly different from the experimentally difficult to ascertain values of *p*=0.118 and *x*=6.56 provided in the paper. The above tabulated values in fact are fairly similar to those presented at the beginning of this matter in 2007 by Doiron-Leyraud and colleagues [17a] measuring field-induced oscillations in the Hall coefficient for an underdoped Y-123 sample. In the case of [17a] the corresponding set of values is:

| m | 9 | 10 | 11 | 12 | 13 | & | 1 |
|---|---|---|---|---|---|---|---|
| $1/B_m$ | 0.01629 | 0.01810 | 0.01991 | 0.02172 | 0.02353 | | 0.001810 T$^{-1}$ |
| $B_m$ | 61.38 | 55.25 | 50.23 | 46.04 | 42.50 | | 552.5 T |

,
with the spacing at 0.00181 T$^{-1}$ now supporting $p$=0.107 and $x$=6.535. Very significantly, since it relates to the matter in hand, the fields here specify, as in the Riggs' $C_p$ data (at 1K), minima in the $R_H$ oscillatory trace.

Near 3K the $C_p^{el}$ trace in [103] actually sign-inverts (see lower traces in fig. 2B, and fig. 2C), with the heat capacity then becoming greatest not halfway through a field cycle but at its beginning. Apart from this phenomenon and the higher 'order' m to which the large $C_p$ oscillations are able to be followed, the situation thus far would appear very like what has gone before. However the most notable aspect of the $C_p$ data is that the above oscillatory component to the electronic specific heat (that $\propto T$) arrives on top of an underlying contribution growing monotonically as $\sqrt{H}$ right from $H$=0. This latter is the form of variation to characterize $d$-wave superconductive gapping of a Fermi surface [105], with the pairs first disrupted by the field sitting in the vicinity of the nodal gaps, – i.e. with $d_{x^2-y^2}$ having **k** near the diagonal 45° directions. In the present scenario such pairs propagate perpendicular and parallel to our 2D domain walls. What is most striking is that this $\sqrt{H}$ contribution continues completely unchanged in form and magnitude right through the irreversibility field near 25 tesla at which the vortex array melts and the oscillatory component to $C_p^{el}$ emerges, just as for all previous 'QO' techniques. The melting of the vortex array, as further explored recently in transport and spin susceptibility measurements made on LBCO($p$ = 0.095) by Wen *et al.* [104], brings entry into a resistive condition. The latter of itself does not, however, destroy the pairing, just the coherence of those pairs, so that, for example, Josephson-type effects are lost. Much as when $T>T_c$, so here too the presence of magnetic fields in excess of 25 tesla admits pairing phenomena still such as the extensive precursor Nernst, Seebeck, Hall, diamagnetic and dielectric properties [75,104,106].

It is evident from the assessed rather low magnitude of $\gamma_{o,o}^{el}$ – 1.85 mJ mole$^{-1}$ K$^{-2}$, rising only to 5 mJ mole$^{-1}$ K$^{-2}$ by 45 T – that the Fermionic quasiparticle content in the system never is great enough to relate here to a Fermi surface fully reconstructed into a field-hardened SDW/CDW condition [107] (let alone to the parent state). The gapping is more elementary, established in the disordered, micro-inhomogenous, underdoped, mixed-valent cuprates by proximity to the Mott-Anderson state, and in particular by what is going on in the stripes. As Riggs *et al.* argue, the two key components to $C_p^{el}$ relate not to two independent bits of the system, as with some customary two-phase condition, but to the much more intimate state of domain wall and domain interior. If we look to the domain interior to support the more standard $\sqrt{H}$ component to the overall signal, then the oscillatory component must be an expression of what is afoot at the domain wall, to which, within the present modelling, the charge holes ($^{8}Cu_{III}^{0}$) and their bosonic double-loadings ($^{10}Cu_{III}^{2-}$) have become segregated. Moreover the latter region in the system is that to which the magnetic vortices appear drawn after the vortex lattice melts. In the real space attribution of the oscillations their sourcing has to derive from the changed interaction of the vortices with the charge of the stripes as the magnetic field is advanced beyond $H_{irr}$. The great benefit of these specific heat measurements has been to disclose continuity with what is happening prior to $H_{irr}$, and to associate melting and restructuring of the vortex lattice with what comes beyond – the oscillatory behaviour.

Referring back now to the transport work of [13] in §3 one is able securely to endorse the two-carrier decomposition made there in terms now of charge motion across and along the diagonal stripes, hole-like and electron-like hopping respectively.

*7.2. Clarification of the revised mode of oscillation numerology.*

From figure 12, with its gradient of 2½ flux tubes per relevant unit hole concentration, one would have, by $p = 1$, flux tubes of 2½$\Phi_o$ to every such *active* hole (this being that against which $p$ Is appraised). From what has been said above, this points within each *supercell* appertaining to doping $p'$ to 2½ flux tubes (units of $\Phi_o$) through the one such site in situ there;

i.e. flux (5176 T x $D(p)^2$) ≡ 2½ tubes (of size of $\Phi_o$) to each $p'$ supercell.

For simplicity let us consider now the specific supercell circumstance of figure 6, one which falls on the line in figure 8 and for which the relevant concentration $p' = {}^9/_{9^2} = 0.111$. The supercell hole content, in this case of 9 holes, occupies the 9 yellow ($^{8}Cu_{III}^{0}$) stripe sites within the pair of crossed diagonals of the 9x9 cell. Here, when under a field then 9-fold diminished in strength to 574 T, the one of these 9 hole sites adjudged to afford the natural choice to bear the above 2½$\Phi_o$ flux-tube allotment is that located at the stripe crossing point. This is the circumstance then to define m=1 'head field' $B_1(p={}^1/_9)$. For the m=2 stage, at one half the $B_1({}^1/_9)$ field strength, only every second supercell will stand so decorated, at m=3 just every third supercell, and so on. When we pass here from the m=n to the m=n+1 condition as the applied field lessens, this structuring alteration supplies the ratcheted change in the observed 'QO' signal, attributable to a seesawing of the chemical potential. Just what the actual geometry of the 'dilution' patterning of supercells is going to be adopted is not evident here (see figure 5 in [27] for possible geometries in relation to the case of m = 13, – though note the stepping unit refers now to supercells, not domains). In particular note the above stepping in the signal-generating dilution process automatically will stand proportional to 1/$B$, *not* 1/$B^2$.

We must look next at how far the above assessment is matched by direct knowledge of the geometry of the vortex array, its symmetry and its scaling.

*7.3. Geometry of the vortex array in HTSC material in increasingly strong magnetic fields.*

At low fields (~1 mT) the vortex array can be directly imaged in the electron microscope via Bitter pattern decoration using cobalt smoke, and it takes there the classic hexatic form, provided defect pinning is slight [108]. Using small-angle neutron scattering (SANS), this general form has been demonstrated to persist to ~0.25 T, by which field the inter-vortex spacing, $s$, has been reduced to ~1000 Å, a spacing just beginning to register against the stripe spacings portrayed earlier. In figure 6 for $p = 0.09$ the stripe spacing stands at 25 Å. This signals the interaction of the vortices with the surrounding nanoscale order could now begin to become as significant as the inter-vortex forces that brought about the original hexagonal close-packing. Indeed in fields above 0.25 T it has been determined for several HTSC systems that the hexagonal array starts adjusting towards a square one, while still remaining within the 'vortex solid' state. White *et al.* [109] by means of SANS on $YBa_2Cu_3O_7$ at 2 K report discrete jumps there in the course of compliance of the vortex array to the lattice, these recorded at 2 and 6.7 T. Reibelt *et al.* [110] also for YBCO7 have detected by magnetization and magneto-caloric measurements such an event near 1 T. Gilardi *et al.* [111], pursuing SANS on LSCO $x$=0.17, have made there the additional highly significant observation that, as square geometry is taken on by the vortex array above 0.8 T, the orientation of that square arrangement is not axial but acquires the 45° orientation of the stripes, as can be ascertained from figure 2b of [111] (in line with the current figure 6).

A yet further observation of considerble significance has been reported in the very recent paper from Yuan Li et al. [112] relating to SANS work on $HgBa_2CuO_{4+\delta}$ crystals. In that tetragonal material they find that the vortex array, even while still triangular in form, locks on to either the *a*- or *b*-axis directions in equal proportions to supply a 12-spotted diffraction ring. What next occurs is particularly telling. Beyond 0.2 T this diffraction pattern rapidly fades away in intensity as $1/\sqrt{B}$ to disappear completely by 0.4 T, with no vortex array diffraction signal re-emerging prior to the limiting field available to Li of 4 T. The latter field is, note, still well below the array-melting field, $H_{irr}$. Y. Li *et al.*'s observation would support the idea that all the flux vortices have become withdrawn from the bulk to anchor themselves randomly to the stripes. The reader is able in figure 4 of the paper from L. Li *et al.* on LSCO [113] to view an example of the typical HTSC vortex phase diagram as projected upon the *x-H* plane. In order to access the 'vortex liquid' phase (at which point vortex-antivortex pairings emerge under local phase fluctuations) one needs to proceed for $x = 0.10$ at $T < 1K$ to fields in excess of 25 T – precisely those fields employed now in the 'quantum oscillation' experiments. Under these conditions the vortices are become so compacted by the applied field and so reorganized by the stripes that the very notion of flux vortices requests replacement by one of flux tubes attached to specific composite carriers as mooted above.

Let us proceed now below with further consideration of the case for $p = 9/9^2$, which sits in the middle of the 'QO' regime, and for which we have the rudimentary, frozen representation of figure 6. We shall employ this figure as basis towards examination of how application of a strong magnetic field develops the situation. For the numerically amenable and attainable case of the m=9 oscillation, we find applying figure 14 that the relevant reciprocal field strength here is $1/0.01575$ tesla$^{-1}$; i.e. $B_9(p=0.111)$ is 63.5 tesla. Upon employing (for a square array) $B_9(^1/_9) = 2\frac{1}{2}\Phi_0/s_9(^1/_9)^2$ (with $s$ the intervortex spacing expressed in nanometers and $\Phi_0 = 2067$ tesla nm$^2$) $s_9(^1/_9)^2$ emerges as 81.3 nm$^2$, giving $s_9(^1/_9)$ as 9.02 nm. With $a_o = 3.85$ Å this spacing would amount to 23.4 basic Cu-Cu unit cells. It will be more advantageous in perceiving what is afoot to transfer attention now to the assessed head-field conditions, for which $B_1(^1/_9)$ is 571 T and $s_1(^1/_9)$ becomes 30.08 Å or 7.81$a_o$. Note the latter spacing has in fact passed somewhat below 9$a_o$, a matter now clearly to require explanation.

In **figure** 17 an extended view of figure 6 is presented incorporating several of the 9$a_o$x9$a_o$ supercells. The interior sections of the domains have for simplicity been filled in here, the two tones of blue marking the opposed chirality of their spin ordering. In addition the centre site of the 9$a_o$ supercell has been translated from having a vacant site at the stripe crossing point there to featuring a Cu atom. The hole at the latter Cu site will be that assigned to carry the 2½$\Phi_0$ flux vortex allottment, represented by the large circle. For the (externally assessed) $B_1(^1/_9)$ field of 571 T we have argued above that every such site in the supercell array is so decorated. The spacing of these points would be expected of course to be 9$a_o$ – hence the above surprise that the experimentally indicated value should appear somewhat less than this. We shall proceed, however, by presuming that the vortex spacing is indeed here 9$a_o$ (i.e. one at the centre of each supercell) and assert that the measured/applied field must be somewhat greater than the field as it is experienced by the material internally. This relaxation in the vortex separation from the calculated value of 7.81$a_o$ to the true value of 9$a_o$ then demands that the applied field strength must exceed the internal field strength in the ratio -

$$\frac{B_1^{app}(^1/_9)}{B_1^{int}(^1/_9)} = \frac{9.00^2}{7.81^2} = \frac{81}{61} \quad ; \quad \text{i.e. with } B_1^{app} = 571 \text{ T one has } B_1^{int} = 430 \text{ T}.$$

The above can be rewritten as $B_1^{app}(^1/_9) = B_1^{int}(^1/_9).(1 + ^{20}/_{61})$. If we suppose this reduction of $B^{int}$ below $B^{app}$ is due to a screening process and one that is directly proportional to the sourcing site concentration, $n_\sigma$ (in nm$^{-2}$), then we mayre-write this expression in the form -

$$B_1^{int} = B_1^{app} - 2n_\sigma.\Phi_0 \text{ (the reason for the factor of 2 will be apparent shortly).}$$

We now have $430 = 571 - 2n_\sigma.\Phi_0$,

from which $n_\sigma = 141/2\Phi_0 = 0.0341$ nm$^{-2}$ or 3.41 Å$^{-2}$. If we express this concentration as $N_\sigma/D(p')^2$ in terms of the supercell unit, it yields the ratio $^{52}/_{81}$ for the *fraction* of Cu sites contributing here to the above magnetic screening. What are the other 29 sites involved in?

If the screening response issues from the fermions in the middle of the domains, hard up against Mott insulation, then the sites that now we are addressing have to be those of the stripes (17 sites per supercell) in conjunction with those contiguous sites intimately interactive with the stripes at their crossing points. The latter provide the 12 remaining sites, 4 around the supercell-centre stripe-crossing point and 8 around the cell-corner crossing point. The function of such $d^\theta$ ($^9$Cu$_{II}^0$) sites has throughout my previous works on HTSC [7,8,10,32,71,114,etc.] been presented as being to double-load the active yellow hole sites

$^8Cu_{III}{}^0$ to $^{10}Cu_{III}{}^{2-}$. The latter local-pair bosons are perceived as bearing a resonant (Feshbach) relation with the chemical potential, this set principally by the majority species ($^9Cu_{II}{}^0$) of the blue regions in our two-subsystem mixed-valent material. This key degeneracy is achieved in negative-$U$ fashion. It is ascribable to the $d$-shell-filling attained at the high-valent site in the course of double-loading, and it involves many-body accommodation of the energy level in question to the acquired $p^6d^{10}$ electron complement. The ready interchange of electron pairs possible between these doubly-loaded sites and the domain interior sites is what yields a superconductivity that is global in character, i.e. of one unified $T_c$. This is comparable with the situation under homogeneous BCS/BEC crossover as it obtains in the optically trapped dilute gas systems. However with the HTSC cuprates, the systems, being chemically inhomogeneous, allow the local-pair binding energy and BCS-related gap energies to diverge in the 'underdoped' regime below $p_c = 0.185$, the former going up, the latter coming down. It is in this underdoped regime, wherein the two subsystems are disengaging, that the presently challenged 'quantum oscillation' experiments are being conducted. The localizing circumstances mean the $d^9$-loaded coordination units of the blue region consolidate their fermionic character and the $d^8$- and $d^{10}$-loaded units, likewise, their bosonic character. The nearer to $p_c$ the less well defined the quantum statistics will become.

As particle exchange is lost between the domain interior and the stripes at low temperatures (and high fields), promoted by the inherent disorder of the mixed-valent chemical system, all 62 *sites* within the blue region will take onboard just one electron. If read as the filling of an entire band of fermion states, as though the situation were homogeneous, this amounts to the perceived fermionic filling factor of $^1/_2 \cdot {}^{62}/_{81}$ - or just short of 0.4.

It has to be recognized that the above electrons, although indeed fermions, are far from being standard Fermi-Landau quasiparticles holding weak mutual interaction. Yet these are the electrons purportedly to uphold standard quantum oscillation behaviour, pursuant upon Fermi surface nesting and charge density wave reconstruction of the FS geometry.

*7.4 Sebastian et al.'s recent conveyance [115/6] of the fault lines in the QO scenario.*

In attempting to consolidate a Fermi-Landau quasiparticle description of the QO data, Sebastian *et al.* in [115] have made a very extensive study of the temperature dependence of the results, and demonstrate them to adhere very closely to a Fermi-Dirac prescription of the DOS in the vicinity of $E_F$. Hence what oscillates near $E_F$ is electron dominated. However this is by no means synonymous with those electrons being Fermi-Landau quasiparticles. The result however for Sebastian *et al.* precludes any further attention to FQHE physics, following the perception that the quantum statistics there is always complex. Similar rejection follows of any circumstance heavily involved with the lattice, as with Alexandrov's polaron/bipolaron conjecture [117]. However the latter has not prevented the authors of [115,25b] turning to CDW reconstruction of the F.S., this in spite of their sensitive analysis failing to reveal any such thing (no change in $g_o$). As we have witnessed, striping and FQHE behaviour in the underdoped cuprates are in fact largely transient and electronic – electronic behaviour that is a highly correlated, many-body one among the fermions with relatively little participation by the lattice, unlike for a standard CDW/PSD.

The above confusion is added to yet further in the latest detailed release from Sebastian *et al.* [116]. The preceeding analysis was accomplished by examining the evolution as a function of temperature of the 'even' moments of the precise waveform specific to the fundamental component in the oscillatory behaviour, the *T* behaviour of which was meticulously tracked between 0.1 and 18 K for fields of betweeen 30 and 55 T. The new analysis now in [116,25b] turns to look at the comparative behaviour between the fundamental component and the second harmonic as extracted from the experimental trace, this here secured over a wider range of field, though more limited temperature range, and accumulating data now for sample tilt angles $\theta$ of up to 60°.

What emerges from this remarkable body of data is the striking result that, for all $\theta$, the amplitude of the second harmonic varies simply as the square of the fundamental frequency, and that the two component signals remain strictly in phase, even through $\theta = 54°$ where the fundamental manifests a 'spin-zero'. The two components clearly monitor the same physics. Such a result is not what is to be expected of a standard Fermi liquid, wherein the chemical potential is effectively fixed within the field-oscillation signal-production by the large reservoir of states in the general vicinity of $E_F$. The latter makes for closer to sinusoidal form in the variation of the amplitude. The new observation, by contrast, supports a circumstance in which signal generation proceeds in a strikingly saw-tooth manner, with the chemical potential not being so buffered, but itself oscillating markedly as the applied field is altered. It is the sort of change to be expected for the scenario advocated earlier, as the particles of the striping features transit from or to the flux lattice 'footprint' in markedly discontinuous fashion as the field is ramped up or down. This transfer impinges on fundamental and second harmonic alike, yielding common angular response. It is not too remarkable then that the effective mass variance of the second harmonic with *T/B* is found not to be of classic form, its effective mass emerging in fact as 25% *lower* than that for the fundamental.

The detailed analysis which Sebastian *et al.* provide permits one additionally to assert that the oscillation source contributing here to the overall signal constitutes the one and only element: there is no further 'complementary' segment as in a 'bipolar' hole and electron Fermi surface reconstruction scenario. Riggs *et al.'s* specific heat data [103], discussed earlier in §7.1, strongly suggested this, and indeed too that the segment of F.S. sensed was nodal in its disposition in the zone, this in keeping with the Fermi surface arcs prescription of photoemission devination. That precisely the same oscillatory periods now materialize in the Seebeck and Hall coefficient responses [74,97], despite the latter coefficients being negatively signed, is not a problem here. In the present understanding of how these oscillations arise, all the carriers respond in the same fashion as the chemical potential ratchets too and fro with *B*, so bringing a common oscillation frequency, dictated at each $p'$ by the FQHE physics lying behind what is portrayed in figures 12-14.

Accordingly then we have need to address what it is about the underdoped mixed-valent HTSC systems and FQHE theory which sets up this encounter between the two hot topics of highly correlated solid state physics. One will need to recall here that the physics behind fractional QHE behaviour is quite distinct from that behind integral QHE behaviour.

*7.5. The nature of the FQHE state.*

Quantum Hall physics and particularly fractional quantum Hall physics is the preserve of exotic interactions [118] and the playground of high-level theorists, which hitherto I have neither felt drawn to, nor am well-equiped to stray. However it has become clear of late that the topic has begun to impinge on the realm of materials far beyond semiconductors and quantum dot edge-states, and in particular on that of non-standard superconductors. When the paper from Sato and Fujimoto [119] was published in November 2010, with its uncompromising title 'Existence of Majorana Fermions and Topological Order in Nodal (*d-wave*) Superconductors with Spin-Orbit Interaction in Exterrnal Magnetic Fields', it was no longer possible shun the topic. This was even closer to the HTSC matter than had been Yang and Zhai's 2008 paper [120] entitled 'Quantum Hall Transition near a Feshbach Resonance in a Rotating Trap'. In November 2010 there came too Ferguson and Goldart's paper [121] on time reversal symmetry (TRS) breaking and the *p*-wave superconductivity in $Sr_2RuO_4$, with discussion there of the existence of domain walls between regions of opposing chirality, of chiral charge currents, and of the appearance of non-integral fluxons at domain wall bends. The conjunction of attractive interactions, the absence of TRS, the vortex liquid state, the Feshbach resonance, BCS/BEC crossover, and the quantum Hall state recommended likewise the 2009 paper from Nikolić [122]. However, because of the details of just what abbreviated modelling was being carried through each time in these papers, the theoretical outcome never appeared clear-cut nor of specific relevance to the HTSC cuprate materials. Where the problem came more sharply into focus was with the deliberations of the current paper. Clearly the FQHE situation for filling factor $v = {}^2/_5$ was the point at which to direct one's attention, and to track down how the details for such a specific case might uphold what was being encountered in the underdoped, mixed-valent cuprates. This meant attempting to make contact with Milanović and Papić's 2010 paper [123] ('The transition from two-component 332 Halperin state to one-component Jain state at filling factor $v = {}^2/_5$'), and especially with its 2007 predecessor from Simon, Rezayi, Cooper and Berdnikov [124] ('Construction of a paired wavefunction for spinless electrons at filling fraction $v = {}^2/_5$'). It is there in particular that the nature of the proposed wavefunction for the 'Gaffnian' state, intermediate between at low *p* the more highly correlated Haldane 332 state and at somewhat higher *p* the Jain heirarchy state, would now look so appropriate to what is being met with in the underdoped HTSC cuprates. Those characteristics are the highly local (negative-*U*) interactions, the well-developed two-subsystem nature of the environment – not just geometrically with regard to the stripes, but also the extant boson/fermion, BEC/BCS crossover situation – this all at a time of enhanced two-dimensionality and proximity to the Mott-Anderson transition. As hole doping is transiting $p = 1/9$ we know there is occurring rapid modification in the tunnelling characteristics, not only between domains and stripes, but also in the third dimension. One of the features of the theoretical analysis presented in [124] is its greater simplicity and directness for flat toroidal geometry, and that is afforded too by the current square 2D striping loops.

What then do Simon *et al.* [124] emerge with as appropriate many-body wavefunction for the $\nu = 2/5$ situation?

While a great many FQHE filling factors relate to states covered by the so-called hierarchy or Jain state wavefunctions, requiring Abelian statistics, some like the $\nu = 1/2$ state are more exotic. For the latter well-known case the Moore-Read Pfaffian ground state is appropriate. The wavefunctions are of a type pioneered by Laughlin, now generalized to the new circumstance regarding the summed angular momentum of the particles, fermions and bosons, that contribute towards forming the ground state. The particles moreover become composite in the presence of added magnetic flux. Such states become labelled by the relevant projection operators, employed here as Hamiltonians, and that relate to the specific character and content of the state. The Pfaffian state bears the label $P_3^2$, the Haffnian for $\nu = 1/3$ with fermions bears $P_3^4$, and sandwiched between these at $\nu = 2/5$ comes the Gaffnian, with label $P_3^3$. These latter states are notably associated with partial particle pairing and with three-body interactions at a point. The paper examines how the Gaffnian wavefunction can be connected through as a 'correlator' to conformal field theory (CFT). The Moore-Read Pfaffian is described in this way by the Ising or $M(4,3)$ minimal model. For the more generalized Gaffnian state we must proceed to $M(5,3)$, which uniquely covers the proposed state. It emerges that CFT $M(5,3)$ is non-unitary. This would indicate the state not to be a true phase but from its further properties to represent a crossover quantum critical range between two flanking states of high similarity ('overlap'), the one to high *p* being the Jain state. To lower *p* the state appears to be a Read-Rezayi state or the Haldane state 332. All these states are associated with pairing, the Jain state being more BCS-like, the lower *p* state more local-pair-like, it being dictated by shorter range interaction than Coulombic. The Gaffnian's intermediary condition is characterized not only by its topology and degeneracy but also by its dynamic character, incorporating the first excited Landau level.

I had at one point thought of including a longer section regarding this highly technical, highly relevant piece of research, but the present paper is already plenty long enough to frighten away readers. The paper from Simon *et al* will have to speak for itself: it calls for being read closely more than once. It is clearly written, well developed and, from my own position, highly comforting, in being highly cited, – it is a very theoretical piece. What from my own perspective appears most significant is that it describes a state that is 3-body in nature and highly dynamic, most suited to describing BEC/BCS crossover — or in my own terms the double-loading charge fluctuations $\quad ^8Cu_{III}^0 + {}^9Cu_{II}^0 + {}^9Cu_{II}^0 \quad \rightarrow \quad ^{10}Cu_{III}^{2-} + {}^8Cu_{II}^{1+} + {}^8Cu_{II}^{1+}$.

**Conclusions**

Accordingly, echoing the manner in which we started out on this present paper, I cannot at all follow the way in which Chang and colleagues have felt obliged in their latest work [126] to address the high-field transport data from underdoped HTSC material. In [13] it was the Hall data on Y-124 that was in question. In the work of LeBoeuf *et al.* [17,97], Chang *et al.* [77b], Daou *et al.* [74] and Laliberté *et al.* [125] it was successively the Hall, Seebeck and Nernst effect results on Y-123 that were dealt with. Now, in the latest release by Chang *et al.* [126] from the same Canadian/French grouping, it is the observed anisotropy of the Nernst effect results which is closely followed, tracked as a function of temperature over the full range below 300 K in magnetic fields of up to 30 T. In the material there being examined (123-YBCO, $\delta$=6.66, $p$=0.12, $T_c$=66 K) the latter field of 30 tesla is sufficient to access the vortex liquid state down to low $T$. The persistent drive witnessed under high fields and low temperatures into a negative signing of $R_H$, $S$ and $\nu$ continues in [126] to be ascribed there to dominant, high mobility, quasiparticles from a small, stripe-phase reconstructed, Fermi surface pocket. In all cases one remains reliant upon the same ubiquitously complicit pocket of high mobility electrons, amounting to but a very small fraction of the parent B.Z. and sporting a Fermi temperature of just 400 K or 30 meV. This outcome is proposed regardless of whether we are dealing with Y-123, or with Y-124, or, as is now acknowledged in the new paper, Eu-substituted LSCO over the $p$ range where the striping is strongly frozen in.

In a crossover, negative-*U*-based, Feshbach resonance the combination of bosons and fermions is one which develops its weighting as a function of temperature and applied field, due to Mott-Anderson localization effects and to the enhancement of striping in a strong magnetic field. The development in geometric segregation in the mixed-valent situation of the doped HTSC systems mounts not only as $p$ falls but also as $H_{app}$ rises. This constitutes a more complex circumstance than the homogeneous BEC/BCS form of crossover encountered in the optically trapped, dilute gas problem. The present strongly two-subsystem environment is one able to carry the superconductivity to the remarkable heights witnessed. It is one that too would take the physics of the correlated fractional quantum Hall effect into a much heightened thermal regime (seemingly 20 K). The filling level imposed by the doping level of close to $\nu = {}^2/_5$ is one very suited to upholding this exotic – potentially very exotic – highly correlated condition. It is a condition strongly emergent in the numerology of the quantum oscillation data itself, and it straightforwardly accounts for why the results are so universal and yet so confined all at the same time. The very small bits of Fermi surface so widely attributed to FS reconstruction under classic CDW behaviour are a myth. The size of the observed frequencies does not derive from *k*-space quantization but from real-space quantization and a simple counting process. That counting is of magnetic vortices on the one hand and of 'holes' into stripes on the other. It is the real-space interaction between these two arrays, vortices and stripes, that leads directly to the oscillatory signals. There are no highly mobile, weakly interacting, Fermi-Landau quasiparticles in these systems at low temperature and high fields.

The signs of the carriers are not dictated by Fermi surface curvature and band masses. The interplay is between coherent and incoherent states, between condensed and uncondensed pairs, and between pairs that partake of local double-loading and those induced pairs that are more BCS-like in nature. The results of [13], where we started, constitute a prime example of the outworking of this complexity.

Only in the narrow range between $p \sim 0.090$ and $0.135$ is the level of metallicity vis-à-vis the Mott-Anderson transition suited to sustaining the fractional quantum Hall condition outlined above. To either end of this range the quantum oscillatory behaviour simply fades away in amplitude, whilst still holding to an oscillatory period rigidly predetermined by figure 14. These periods are ones conceived directly in real space and are not the product of some Fermi surface nesting, $k$-space-dictated process. What has been described is most appropriate to the BEC/BCS crossover understanding of cuprate HTSC behaviour.

**Acknowledgements**

My continued thanks go to members of the HTSC grouping in Bristol for enduring my line of attack on HTSC matters well past retirement, and to Prof. Nigel Hussey in particular for permitting me to see and play around with their data prior to publication. I would like too to thank many of the students within the group for keeping me young and argumentative, in particular Mr Chris Lester. Thanks finally are due to the University for a recent extension of my SRF appointment to 2013.

**TABLE 1   (for p24)**     Definition of terms

- $p = N(p)/D(p)^2$ .                    ...... (1)   $N(p)$, hole content in 'square supercell'.
  $D(p)$, 'supercell' edge in basic units $a_o$ (nm)
  $p$, fraction within supercell of unit cells
  bearing holes.

- $B(p) = \phi(p)/A(p)$ ,                ...... (2)

  $\equiv M(p).\Phi_o/(n.D(p)^2)$.       ...... (3)   $M(p)$, no. of fluxons through $n$ supercells
  (standard fluxon $\Phi_o = h/2e = 2067$ T nm$^2$).

- Divide (3) by (1) -

  $$\frac{B(p)}{p} = \frac{M(p)}{n.N(p)} . \Phi_o .$$       ...... (4)

- With  $M(p)/n = \mu(p)$                ...... (5)   Number of fluxons through *one* 'supercell'.
  reach
  $$\frac{B(p)}{p} = \frac{\mu(p).\Phi_o}{N(p)} .$$         ...... (6)   (- just standard definitions so far).

- But *experimentally* $B_1(p) = c_1.p$ . ...... (7)   $B_1(p)$ being characterizing 'head fields' to
  sequences $B_m(p) = B_1(p)/m$  (m integer).

- Comparing (7) with (6) we have for all $p$

  $$c_1 = \frac{\mu_1(p).\Phi_o}{N(p)} ,$$                  ...... (8)   making $\mu_1(p')/N(p') = \kappa'$ for any $p'$.

- Need to evaluate $\mu_1(p)$
  where  $\mu_1(p) = M_1(p)/n$           (- $\mu(p)$ some rational fraction) .

- **Is $\mu_1(p') = 1$ under any set $p'$ ?**    — as was taken in [27]   (- so making $\kappa' = 1/N(p')$),

  i.e. all head fields $B_1(p')$ would see the situation  'one fluxon through one supercell'.

- Then at *rational* fractional fields   $B_m(p') \equiv B_1(p')/m$    (m – running integer),

  one would have    $B_m(p') = 1/m .(\Phi_o/D(p')^2)$  ,

  e.g. at field m = 9 would have to go to 9 'supercells' to embrace one fluxon,
  – the 'fluxon footprint' $F(p') = 9$.

- Free energy ratchet sequence proceeds then as $B_1(p')/m$
  with  $B_1(p') = \Phi_o/D(p')^2$.

- The above (besides $\kappa = 1/N$) presumes standard fluxons = $\Phi_o$ , rather than $\Phi_o/2$, etc.,
  - and also note 'supercell' area $D(p)^2$ is not as yet actually specified.

- Questions arising regarding this original procedure in [27]:
  Relating to eq$^n$ 2  (i) Is true unit flux here indeed $\Phi_o$ ?
  Relating to eq$^n$ 1  (ii) What is to be used for 'supercell' unit area $D(p)^2$ ?
  Relating to eq$^n$ 8  (iii) Is $\mu_1(p)$ really appropriately selected as being 1 ?
  N.B. in (8) a) $\Phi_o$ is explicitly present, and $N(p)$ is very closely related to $D(p)^2$.
  Is $\mu_1(p')$ primarily dictated by $D(p')^2$ or by $N(p')$ ?
  b) Empirically determined value of $c_1$ is 5170 tesla per active centre.
  (see fig.12 for YBCO$_x$, but $c_1$ probably same for all HTSC systems.)

**Index of first authors**

**Figure Captions**

**Figure 1.** (p6)

A full collection is presented in [13] of (a) $\rho_{xx}(H,T)$ magnetoresistance and (b) $\rho_{xy}(H,T)$ Hall data from the underdoped compound $YBa_2Cu_4O_8$ obtained by Rourke *et al* in fields of up to 60 tesla for temperatures below 70 K. The Hall data have been reproduced here with slight modification. Even at very low temperatures bulk superconductivity is lost above 40 tesla. The most striking feature observed in the high-field, low-temperature region is that the material's Hall resisitvity has become negative, as against the positive sign of standard conditions. Between 20 K and 40 K this crossover is seen in progress. Such a sign change has been widely reported for other underdoped, mixed-valent HTSC materials in which standard Fermi liquid behaviour is being surrendered under the resonant B-F scattering and encroaching Mott localization. [Reproduced with permission of APS Publishing]

**Figure 2.** (p6)

Standard and Hall planar resistivities $\rho_{xx}$ and $\rho_{xy}$ for YBCO-124 as functions of applied magnetic field at three selected temperatures 20, 30 and 40 K, from Rourke *et al* [13]. The vertical arrows mark the point of entry into the $\rho_{xx}$-linear *high-field* regime at each temperature. The dashed lines indicate extrapolations back to $T = 0$ K made in accord with the two semiclassical equations given in the text, for this pre-superconductive (mixed state) underdoped behaviour. [Reproduced with permission of APS Publishing]

**Figure 3.** (p6)

The decomposition performed by Rourke *et al* in [13] into positively and negatively signed charge components to $\sigma$ and $R_H$ deemed operative for the underdoped 124 material, executed as a function of $T$ and relating to the pre-superconductive, high-field regime. These components are simultaneously constrained to satisfy both the $R_H(T)$ and $\sigma(T)$ equations given in the text. [Reproduced with permission of APS Publishing]

**Figure 4.** (p10)          (colour on line)

Frozen representation of charge stripes and spin domains in the $CuO_2$ planes for an idealized circumstance with supercell wavevector $\delta = {}^1/_8$ and hole count $p = {}^1/_8$, when allotted diagonal 2D geometry as in [32]. The small dots mark the location of the copper atoms and the colouring marks the enclosing oxygen coordination units. The $d^8$ holes ($^8Cu_{III}^0$ - in yellow/light grey) are shown segregated into the stripes, and alternate there with spin-free $d^9$ sites ($^9Cu_{II}^0$ - in green/mid-grey). The bulk of $d^9$ coordination units freezing into the domain interiors are shown in blue/dark grey. The circulatory spin patterns manifested within the latter are as far as possible antiferromagnetic in alignment, and take on opposite overall chirality between neighbouring domains. Regarding this spin array the stripe directions constitute a mirror, but the primary *x* and *y* axes do not. This time reversal symmetry breaking away from the simple, non-striped, parent structure has in [33] been held responsible for the magnetic circular dichroism detected in [52] and later works. The $^8Cu_{III}^0$ hole sites are host to the bosonic double-loadings $^{10}Cu_{III}^{2-}$, these perceived as being near-resonant with $E_F$ [7,ff].

**Figure 5.** (p10)     (colour on line)

Similar representation to that employed in figure 4 now made for the parameters $\delta = {}^1/_7$, $p = {}^8/_{7^2}$ (i.e. 0.1632, just beyond $T_c^{max}(p)$). Note now that the arrangements of both holes and spins around the supercell's centre and corners differ here, unlike in figure 4. In each figure, however, the phase break in the antiferromagnetic spin alignment along the primary axes between domains amounts to $a_o$.

**Figure 6.** (p11)     (colour on line)

Similar representation to those in figs. 4 and 5, but made now for $\delta = {}^1/_9$, $p = {}^9/_{9^2}$, both $\delta$ and $p$ taking here the same value, as with fig.4. In figure 6 though, as in fig. 5, the arrangements around cell centre and cell corners are distinct. Also as in fig. 5, for such odd-number-edged supercells their domain centres do not fall at a copper atom.

**Figure 7.** (p16)     (colour on line)

**a)** Stylized conception of jogging in supercell wall geometry and how this affects diffuseness of resulting diffraction. The chain or $b$ direction in YBCO$_x$ is the direction of prevalent wall periodicity faulting, bringing weak diffuse $b^*$ diffraction spotting [49]. Here defect cells of size (10x9) and (8x9) arise within the regular (9x9) supercell array – with focus here on the areas.

**b)** Revised conception of what jogging accomplishes, with focus now upon hole content of the stripe circuits around the basic and jog-modified domains. The base supercell here is $9a_o \times 9a_o$ in which the stripe domain circuits at $p = {}^9/_{9^2}$ contain 9 holes. The unit jogging illustrated produces circuits then of 8 and of 10 holes in the jogged region of stripe displacement.

**Figure 8.** (p23)

Development of figure 15d from [55] by Hücker *et al* on the striped state of (La$_{2-x}$Ba$_x$)CuO$_4$. The figure shows the monotonic (at this level step-free) increase of the ($H$=0) incommensurability parameter $\delta$ with Ba content $x$ across the range where for this system $T_c$ displays a very marked dip. In LBCO $x$ may be taken with good confidence as closely identifying planar hole concentration $p$. The following features stand out; (i) no lock-in at $p = {}^1/_8$ to $\delta = {}^1/_8$; (ii) $p = {}^1/_9$, $\delta = {}^1/_9$ does sit on the curve; (iii) $p = {}^1/_{10}$, $\delta = {}^1/_{10}$ however does not; (iv) $p = {}^9/_{10^2}$, $\delta = {}^1/_{10}$ though does; (v) $\delta$ exceeds ${}^1/_8$ only beyond $p = 0.1563$. The large circled crosses mark commensurate stripe conditions of the type presented in figures 4-6:

| cross | F | E | D | C | B | A |
|---|---|---|---|---|---|---|
| conc. $p$ | ${}^8/_{7^2}$ | ${}^{10}/_{8^2}$ | ${}^1/_2({}^1/_9+{}^1/_8)$ | ${}^9/_{9^2}$ | ${}^9/_{10^2}$ | ${}^{10}/_{12^2}$ |
| = | 0.1632 | 0.1563† | 0.1181 | 0.1111 | 0.0900 | 0.0694 |
| see | fig.5 | [32,fig.4] | text | fig.6 | fig.9 | ‡ |

† This condition involves five-hole clusters at both cell-corner and cell-centre dc. crossing points. It is feasibly where $T_c$ maximizes [32].

‡ It is possible for the condition to involve similarly constituted five-hole clusters – although now isolated along the stripes.

**Figure 9**. (p24)  (colour on line)

Representation like those made in figs 4 to 6 but now for the situation $\delta = {}^1/_{10}$, $p = {}^9/_{10^2}$. This experimentally favoured array possesses a strictly alternating sequence within the stripes of yellow and green sites, although, note, the cell centres and cell corners again have distinct geometries as in figs. 5 and 6.

**Figure 10**. (p24)

Plot of data assembled in LeBoeuf *et al* [97] for underdoped YBCO$_x$-123 of adduced planar hole concentration *p* vs. oxygen content *x*. As with the related data in figure 11 it is apparent a two-phase situation arises between -O$_{6.560}$ and -O$_{6.666}$ over which the *p* value remains anchored at 0.1180, the mean of ${}^1/_9$ and ${}^1/_8$. The plot gradients are quite different across this break, below which *p* clearly bears regular numeric form, i.e. does not signal any complex fermiological input.

**Figure 11**. (p24)

Plot of data assembled in LeBoeuf *et al* [97] for underdoped YBCO$_x$-123 of $T_c$ vs oxygen content *x*. Again it is apparent a two-phase situation arises between -O$_{6.560}$ and -O$_{6.666}$ over which the *p* value remains anchored at 0.1180, the mean of ${}^1/_9$ and ${}^1/_8$. Both this figure and figure 10 indicate that the low *p* condition starts to be unstable above -O$_{6.500}$ at which *p* is very close to ${}^1/_{11}$. The greatest depression in $T_c$ from simple parabolic form looks to occur at -O$_{6.666}$. $T_c$ finally drops to zero close to -O$_{6.300}$.

**Figure 12**. (p24)

Plot of a series of 'head fields' $B_1(p)$ in tesla for variously underdoped YBCO-123 specimens and YBCO-124 as deduced from the QO experiments. The data have been assembled in a very recent review by Vignolle *et al* [99]. The as-given points have been linked there by the dotted line. My own direct space interpretation of events would steepen this line to the solid one incorporating the origin. The latter line is much supported upon shifting the YBCO-124 point substantially to the left to $p = 0.128$ (see text) and marginally moving the point below from 0.12 to 0.116 (i.e. into the two-phase regime of YBCO-123; see text). The gradient of the new line, from its axial crossing points (0.09,465) and (0.135,700), is seen to take on the very telling value of 2.50 flux tubes per Cu ($p=1$); i.e. $(465/\Phi_o)/0.09 = (700/\Phi_o)/0.135 = 2.50$.

**Figure 13**. (p26)

The linear relation between $B_m(p)$ and *p* for a group of oscillation-specifying integers, m, falling within the experimental field range, under the global constraint that the gradient $\kappa$ of the m=1 line $B_1(p)/\Phi_o$ vs *p* is ${}^5/_2$ for all underdoped HTSC systems, i.e. $c_1 = 5167$ tesla per planar Cu ($p=1$). Here then we find $B_{10}$ at $p = 0.1$ holds a value of 51.67 tesla.

**Figure 14.** (p26)

Reciprocal inversion of the information given in figure 13 to supply $1/B_m(p)$ vs *p* plots for the relevant m as indicated. The specified values of $1/B_m(p)$ given at $p = 0.105$ and 0.128 relate to experimental work on Y-123 and Y-124 and are dealt with in figures 15 and 16 respectively.

**Figure 15.** (p26)

The field-oscillatory magnetic torque data of Jaudet *et al* [18] obtained from a Y-123 sample with *p* near $^1/_{10}$. The vertical lines mark out the period of the oscillation as being 0.001845 $T^{-1}$ and pick out locations close to the trace minima when made to accord with a $1/B_m(p)$ value of 0.01845 at m = 10. This implies, in compliance with figure 14, a 'head field' of 542 T and a planar *p* value of 0.105. The detailed values referring to this trace are

| m | 10 | 11 | 12 | 13 | 14 | 15 | 16 | & 1 |
|---|---|---|---|---|---|---|---|---|
| $1/B_m$ | 0.01845 | 0.02029 | 0.02214 | 0.02398 | 0.02583 | 0.02767 | 0.02952 | 0.001845 $T^{-1}$ |
| $B_m$ | 54.20 | 49.27 | 45.17 | 41.69 | 38.71 | 36.13 | 33.88 | 542 T . |

In this magnetic torque trace the above features relate to $\psi = -\pi/2$ sinusoidal phasing (see text). [Data re-presented with permission of APS Publishing]

**Figure 16.** (p26)

The field-oscillatory inductance signal from Y-124 recorded by Yelland *et al* [16b]. The vertical lines mark out the period as being 0.001515 $T^{-1}$, and identify locations $1/B_m$ close to sine 180° etc., to accord with a value of 0.01515 $T^{-1}$ at m = 10. This fixes the 'head field' as being 660 T and the planar *p* value for Y-124 as 0.1277 so gaining compliance with figure 14 and the modification made in figure 12. Recall from table 1

$$B_1(p')/\Phi_o = (c_1/\Phi_o).p' = (\mu(p')/N(p')).p' = {}^5/_2.p' ,$$

so $p' = {}^{660}/_{2067} \times {}^2/_5 = 0.1277$ .

The detailed values referring to this trace are

| m | 8 | 9 | 10 | 11 | 12 | 13 | 14 | & 1 |
|---|---|---|---|---|---|---|---|---|
| $1/B_m$ | 0.01212 | 0.01363 | 0.01515 | 0.01666 | 0.01818 | 0.01969 | 0.02121 | 0.001515 $T^{-1}$ |
| $B_m$ | 82.51 | 73.37 | 66.00 | 60.02 | 55.00 | 50.79 | 47.15 | 660.2 T . |

In this inductance trace the above $1/B_m$ features relate to $\psi = \pi$ sinusoidal phasing (see text). [Data re-presented with permission of APS Publishing]

**Figure 17.** (p32)   (colour on line)

An applied field of 571 T in conjunction with a flux quantum of ${}^5/_2\Phi_o$ would, with a more standard superconductor, be expected to generate a square vortex array of spacing 30.1 Å. For the present HTSC material, this field becomes screened internally to 430 T, associated for the same ${}^5/_2\Phi_o$ quantum with a flux spacing now of 34.6 Å or $9a_o$. This spacing brings commensuration with the striping supercell; the ${}^5/_2\Phi_o$ flux-tube is taken as being closely associated with the stripe crossing-point coordination unit central to the $9a_o$-by-$9a_o$ supercell as it is drawn here (compare fig.6). The array of such vortices, marked in the figure by large circles, defines the m=1 'head field' condition appertaining to externally appraised field 571 T. By oscillation order number m=9 there remains just a single vortex central to the group of 9 supercells indicated, this occurring at applied field $B_9(^1/_9) = 571/9$ or 63.5 T. The above ${}^5/_2\Phi_o$ flux quantum is associated with FQHE filling factor $\nu = {}^2/_5$.

FIGURE 1

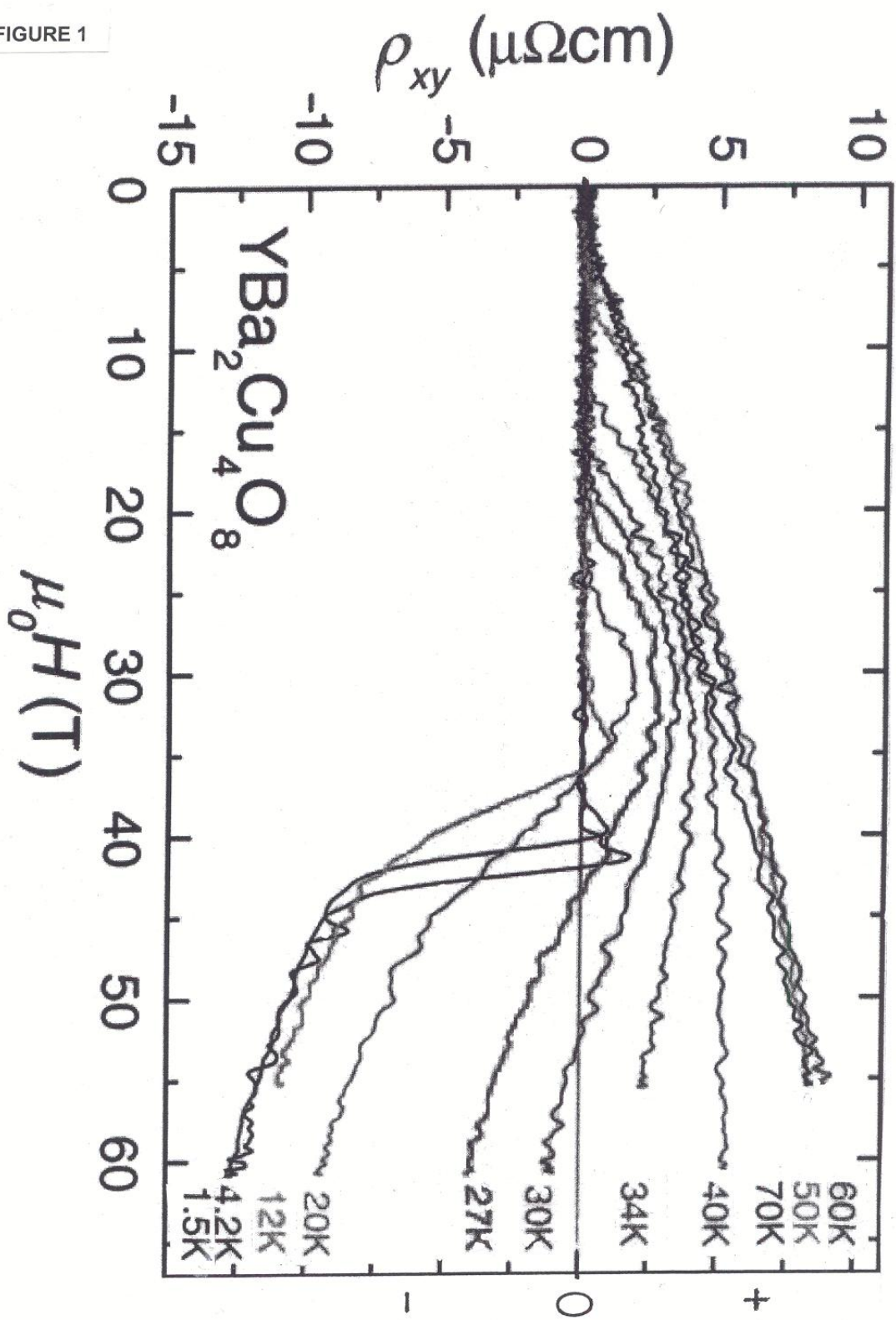

f1

FIGURE 2

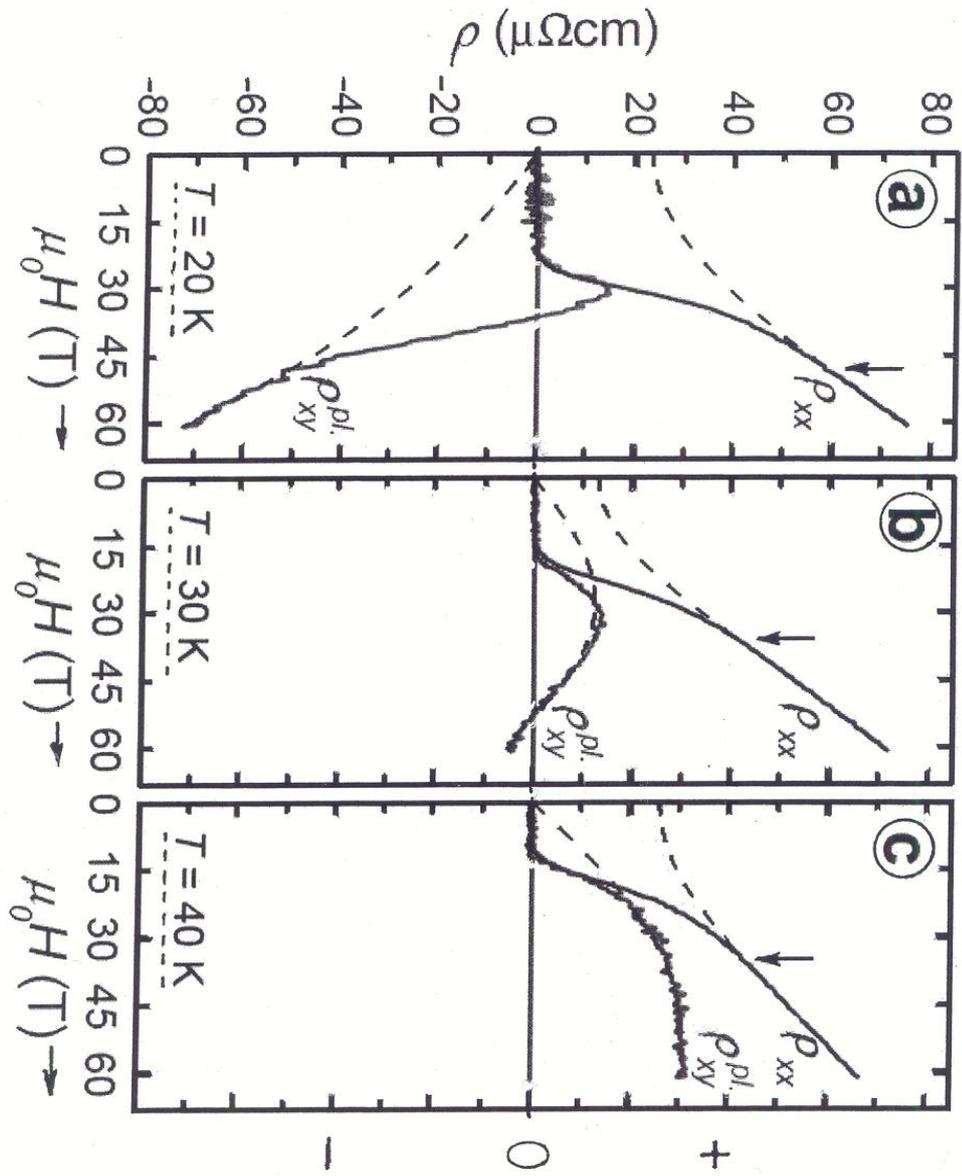

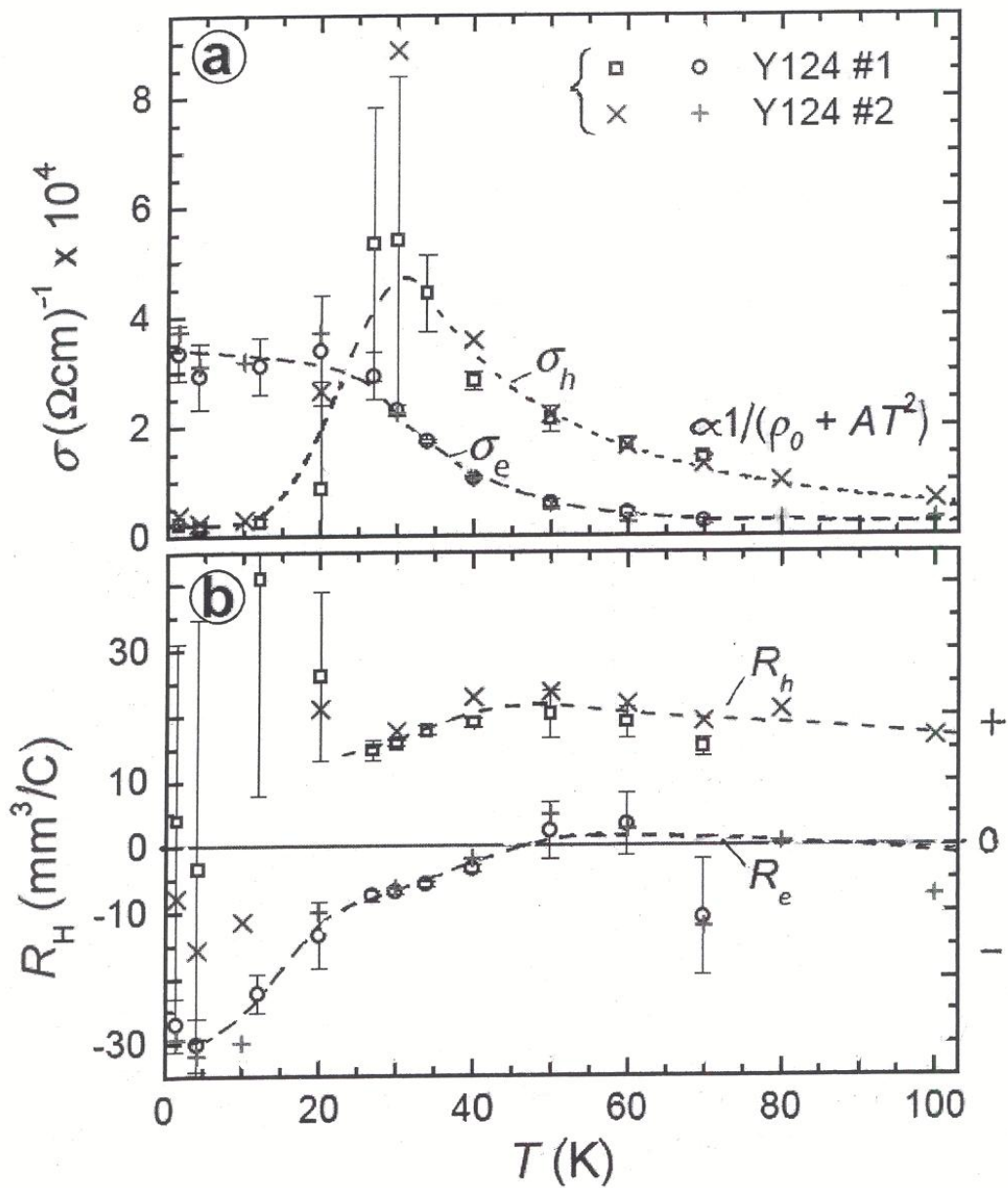

**FIGURE 4**

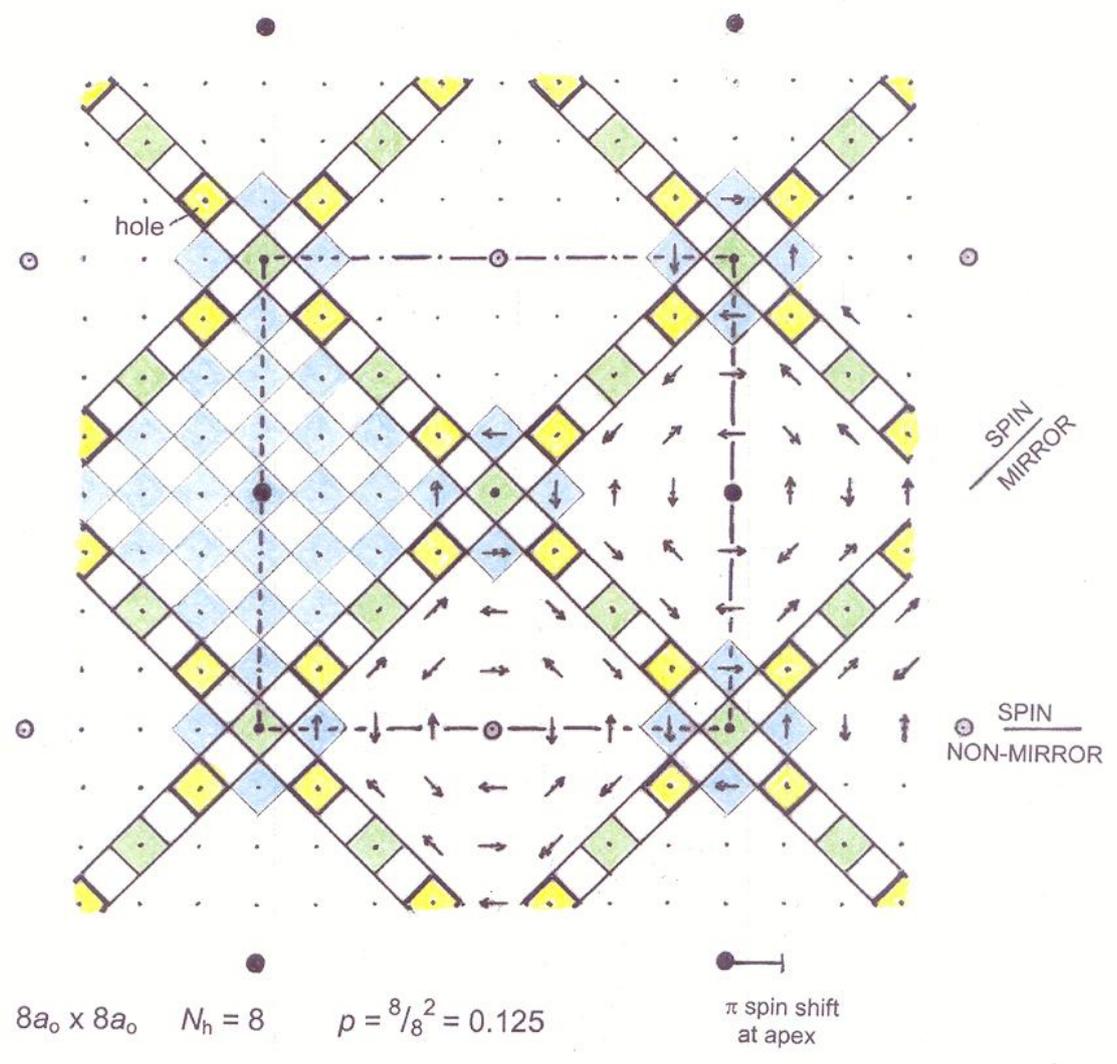

$8a_o \times 8a_o$    $N_h = 8$    $p = {}^8/_8{}^2 = 0.125$    π spin shift at apex

**FIGURE 5**

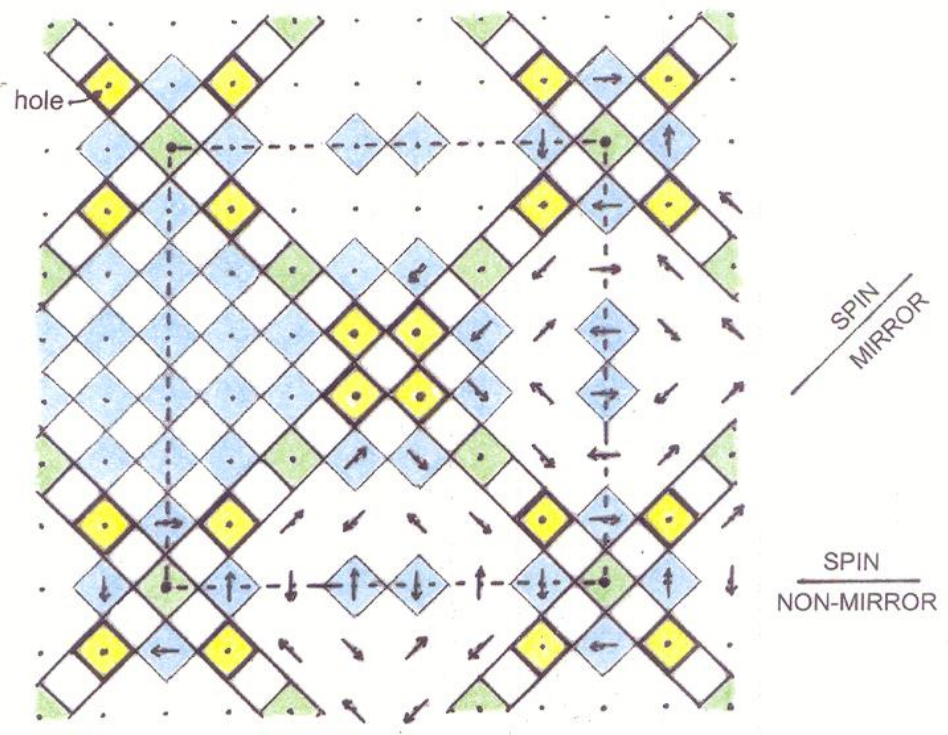

$7a_o \times 7a_o \quad N_h = 8 \quad p = {^8}/{_7}^2 = 0.163$

FIGURE 6

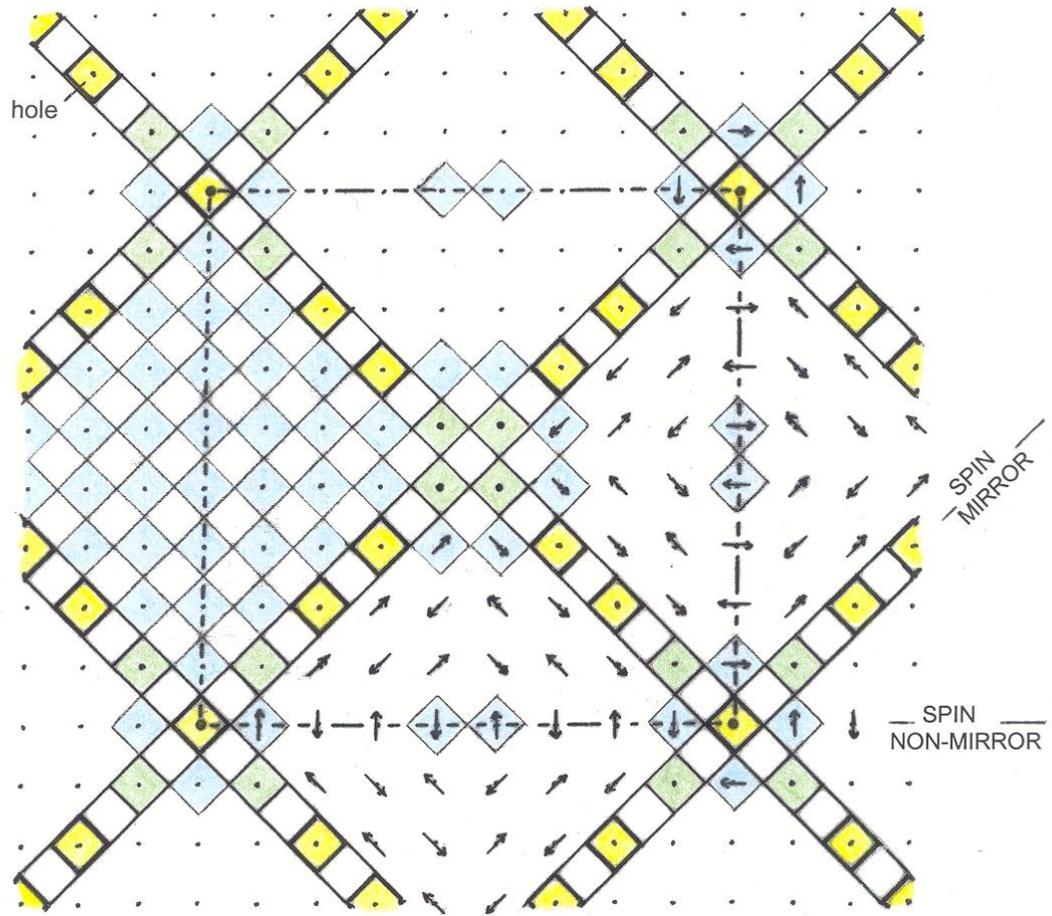

$9a_o \times 9a_o$    $N_h = 9$    $p = {}^9/_{9^2} = 0.111$    $\pi$ spin shift at apex

FIGURE 7a

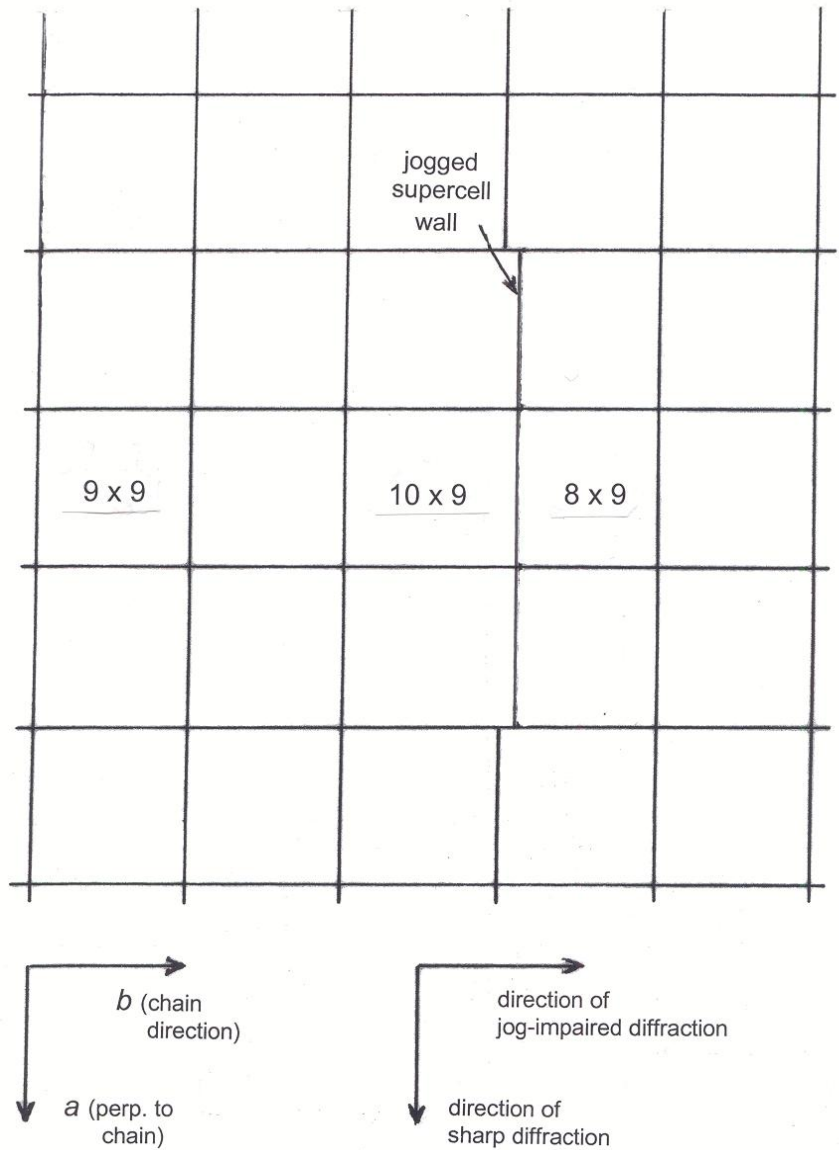

f7a

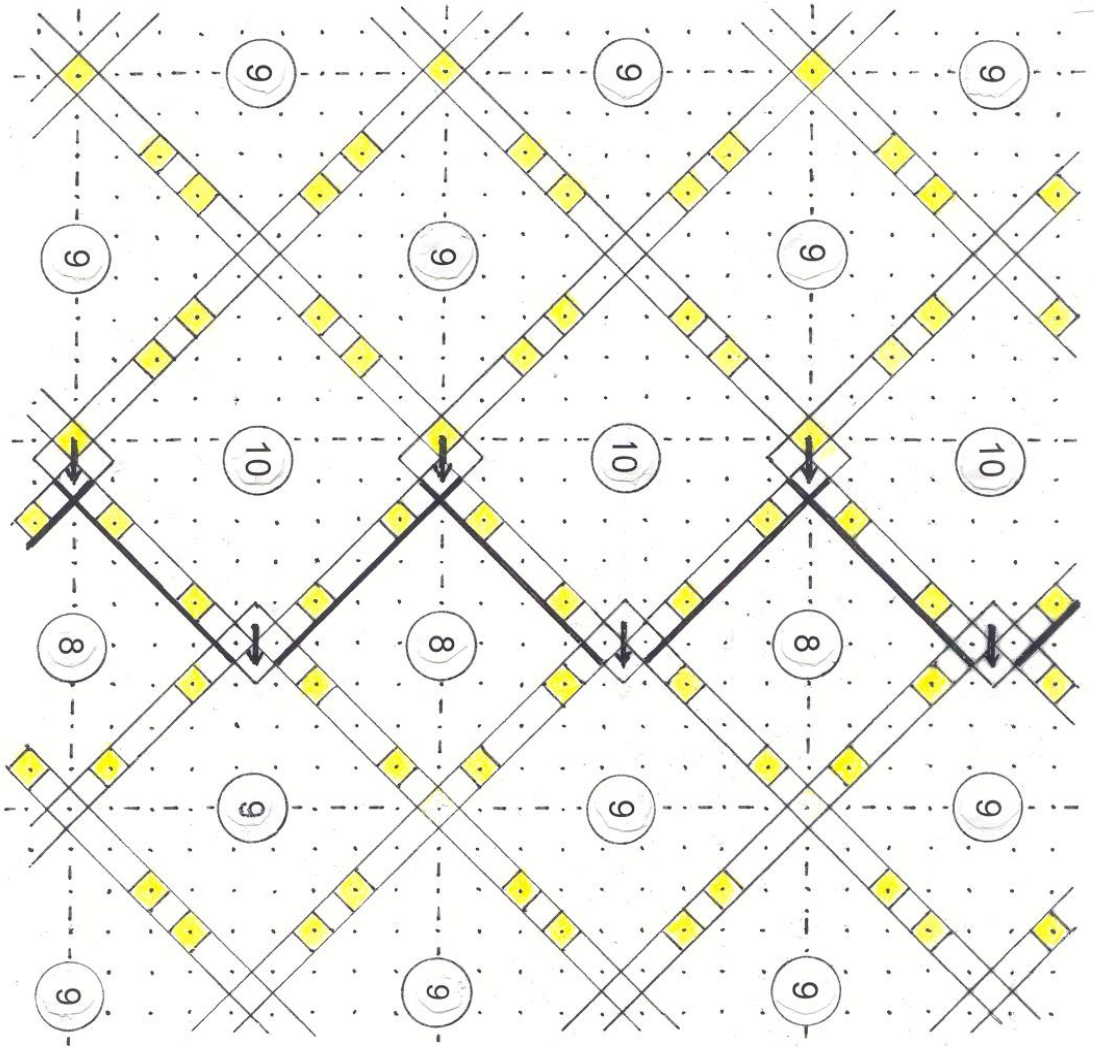

FIGURE 7b

f7b

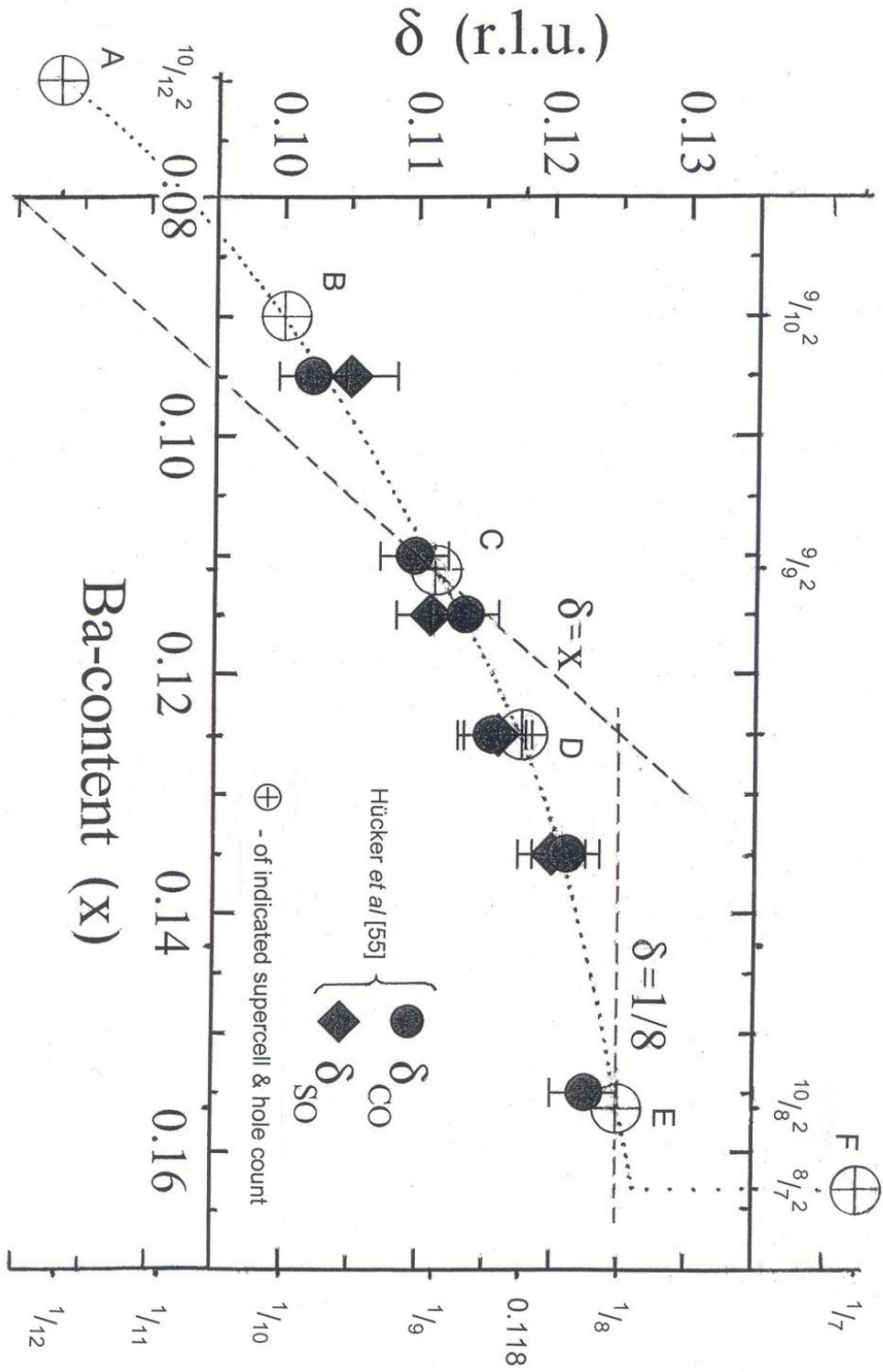

FIGURE 8

f8

FIGURE 9

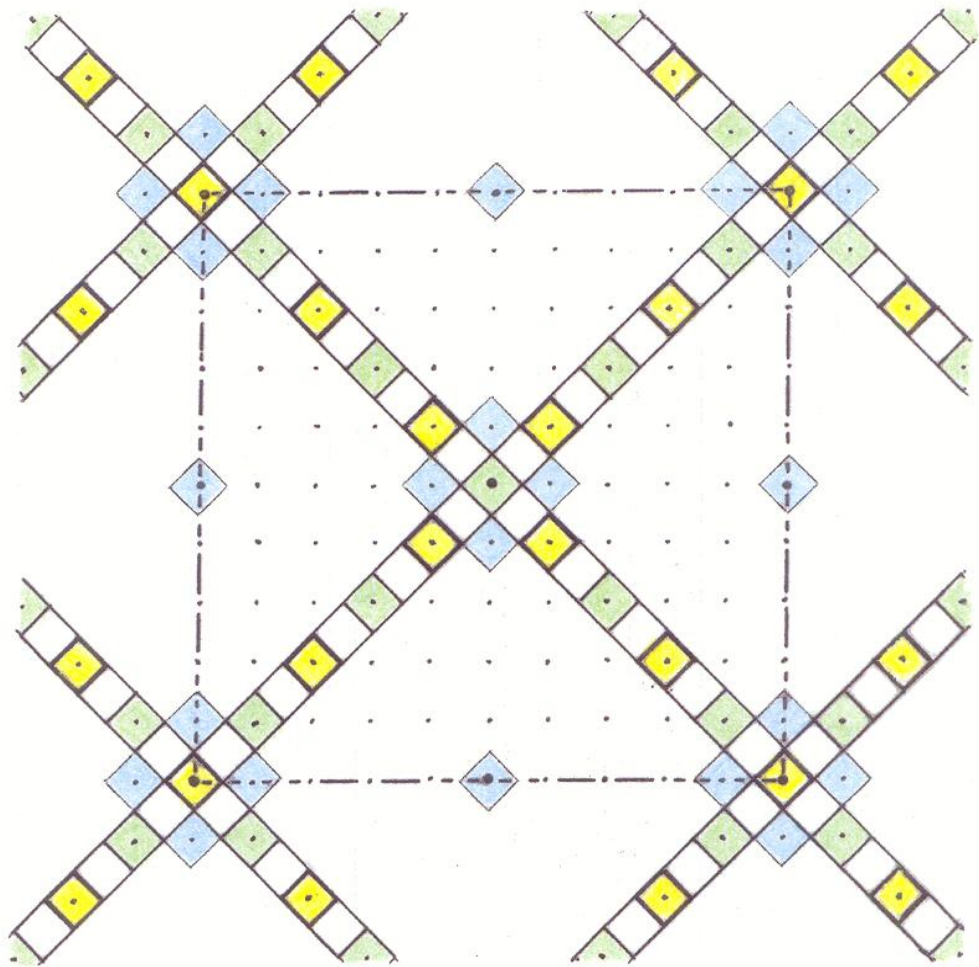

$10a_o \times 10a_o$     $N_h = 9$     $p = {}^9/_{10}{}^2 = 0.090$

FIGURE 10

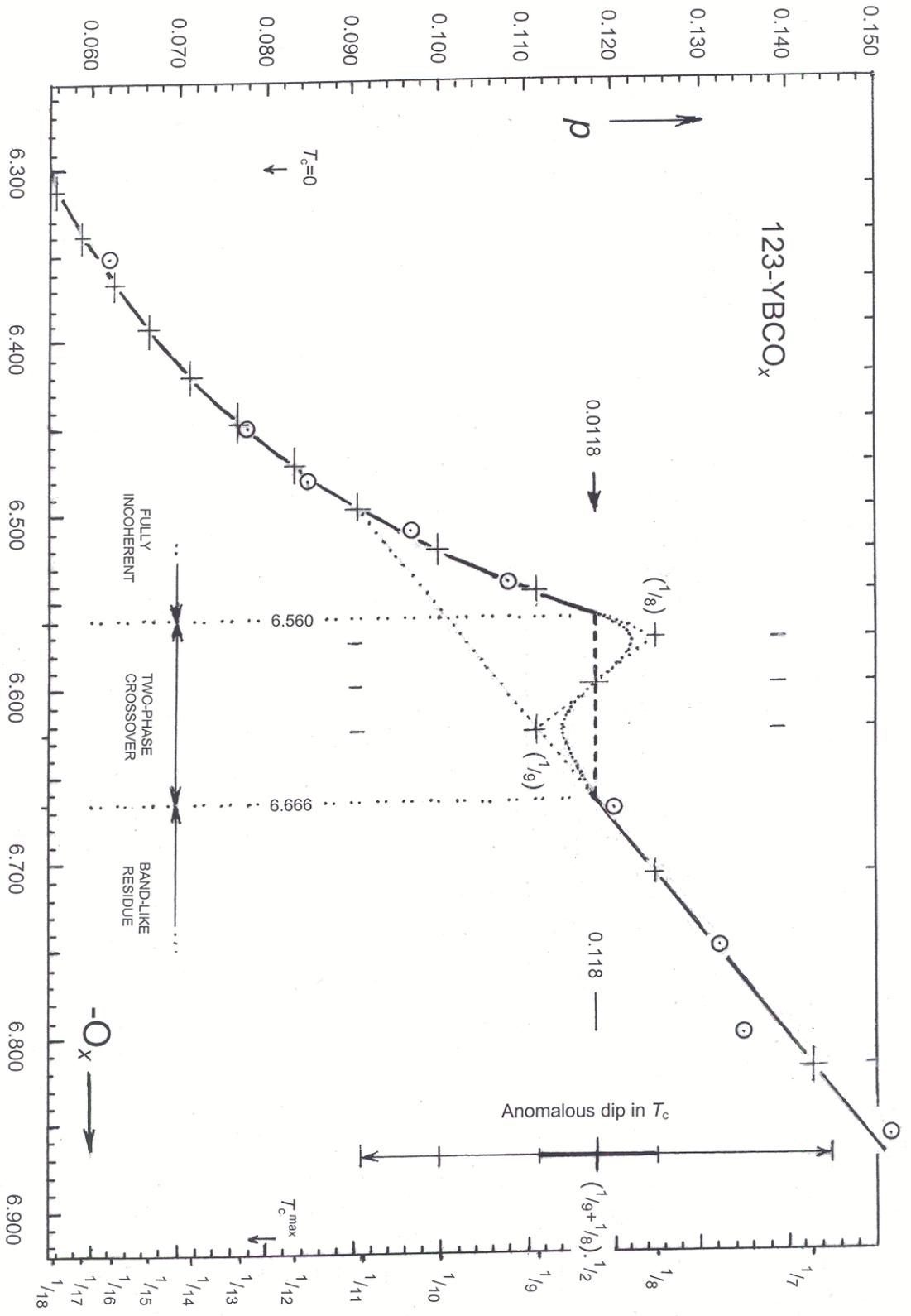

f10

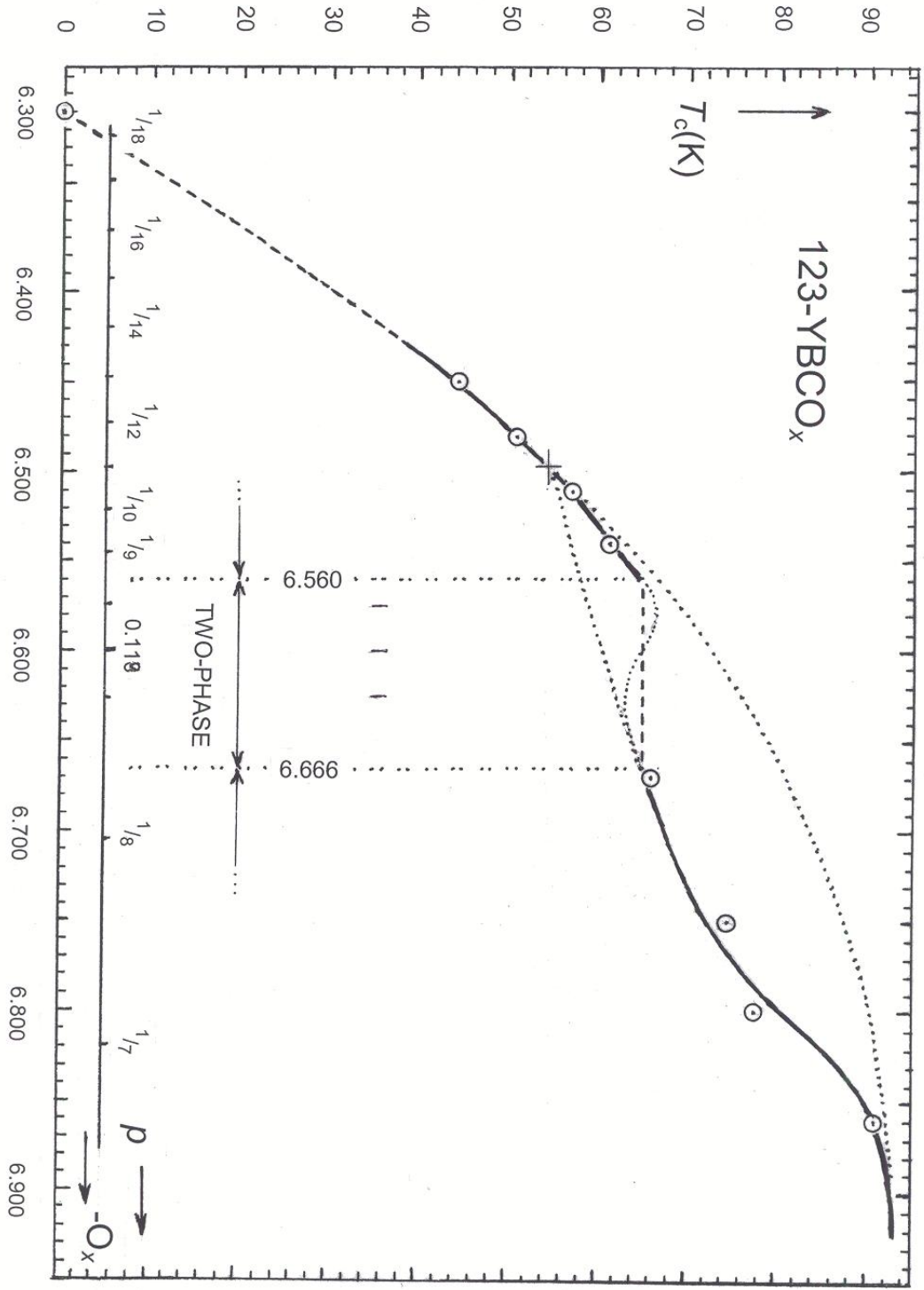

FIGURE 11

f11

FIGURE 12

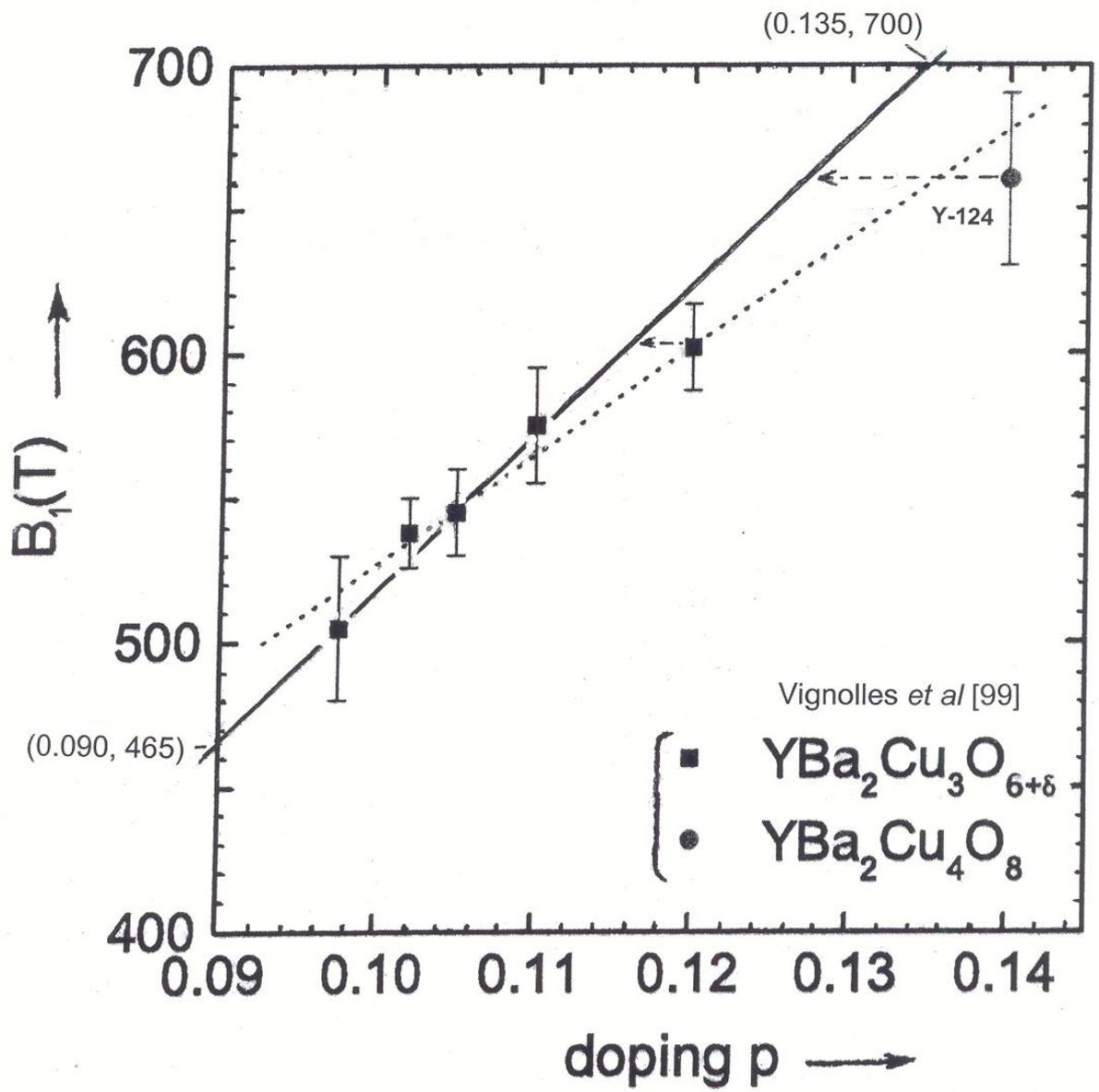

f12

FIGURE 13

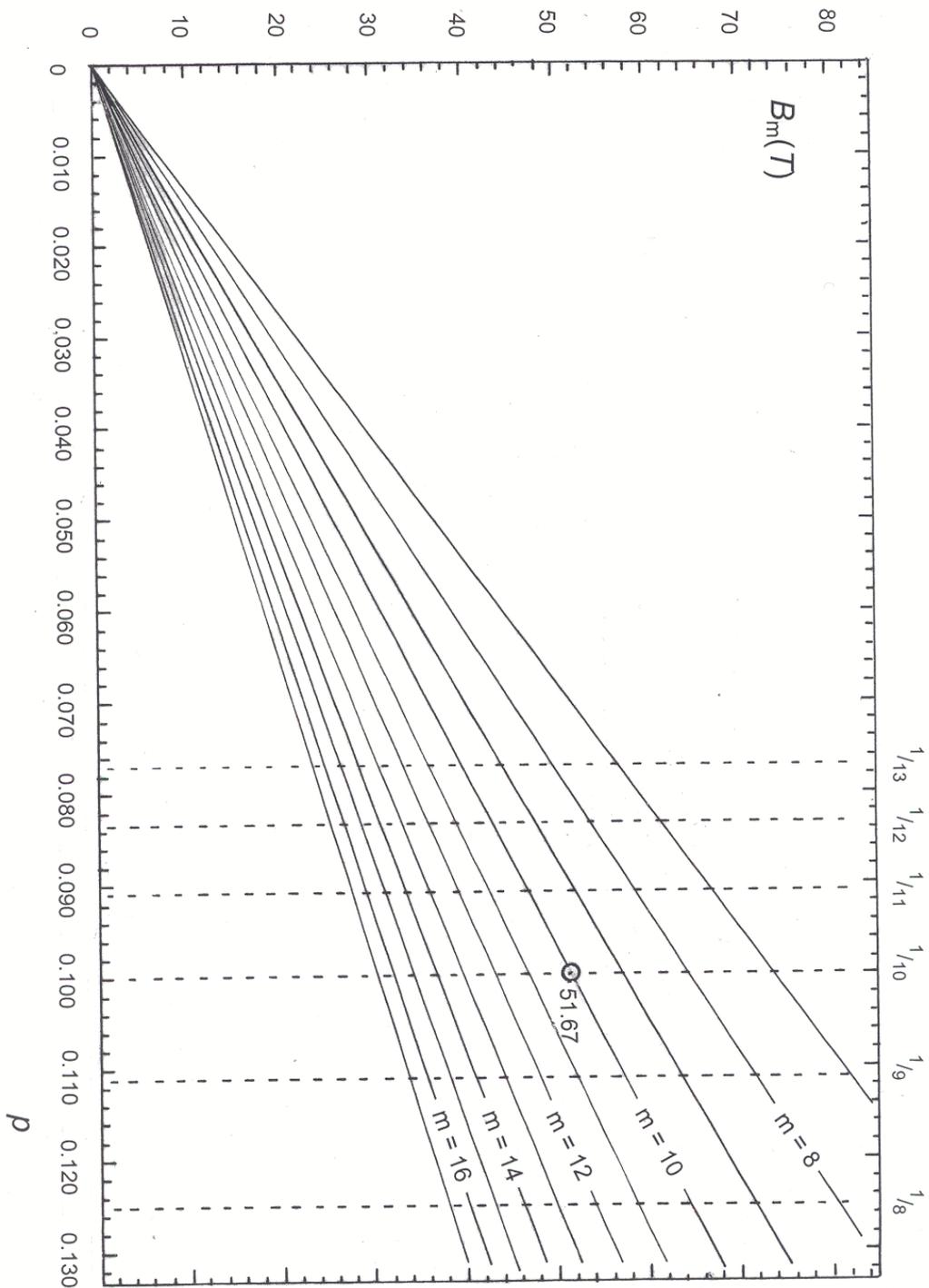

f13

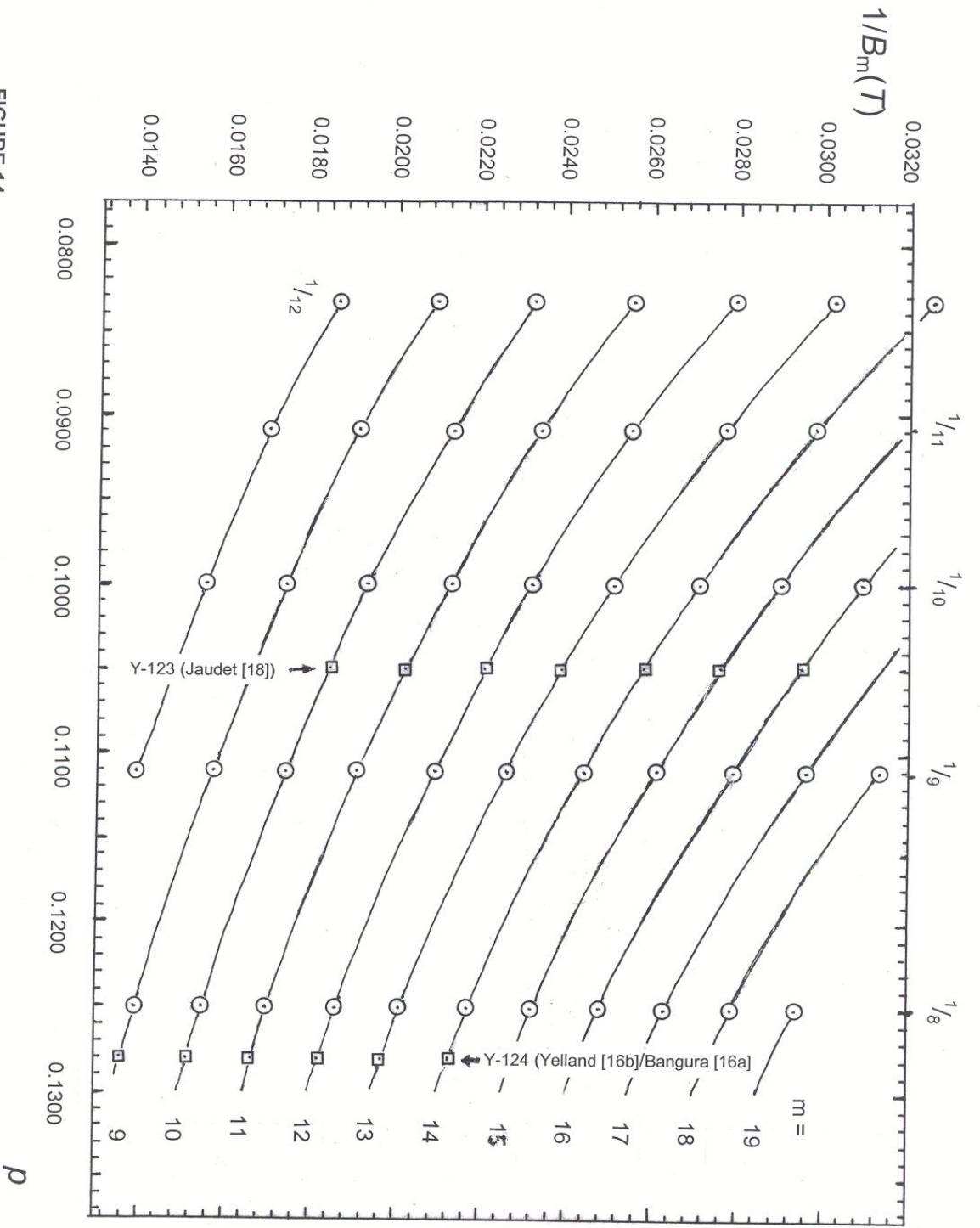

FIGURE 14

FIGURE 15

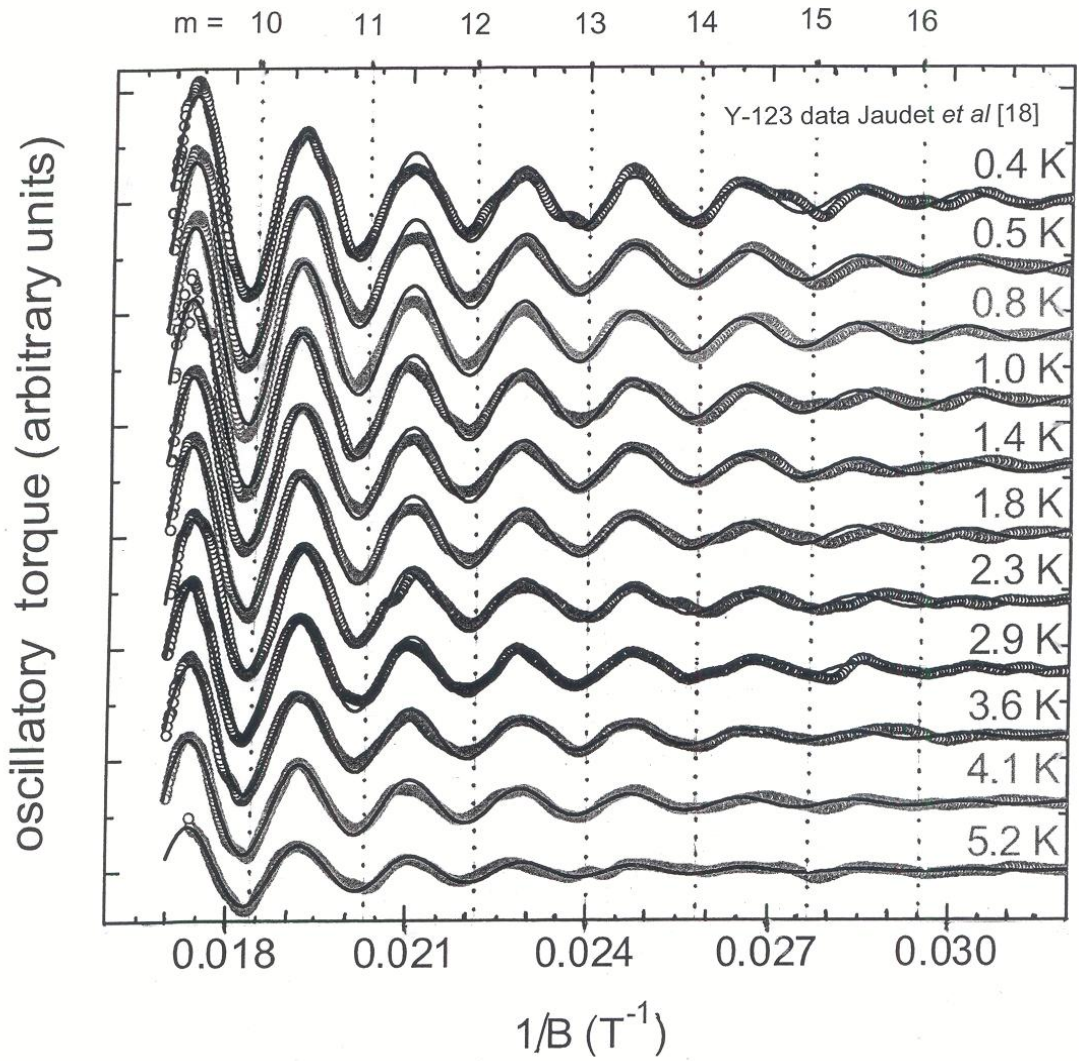

f15

FIGURE 16

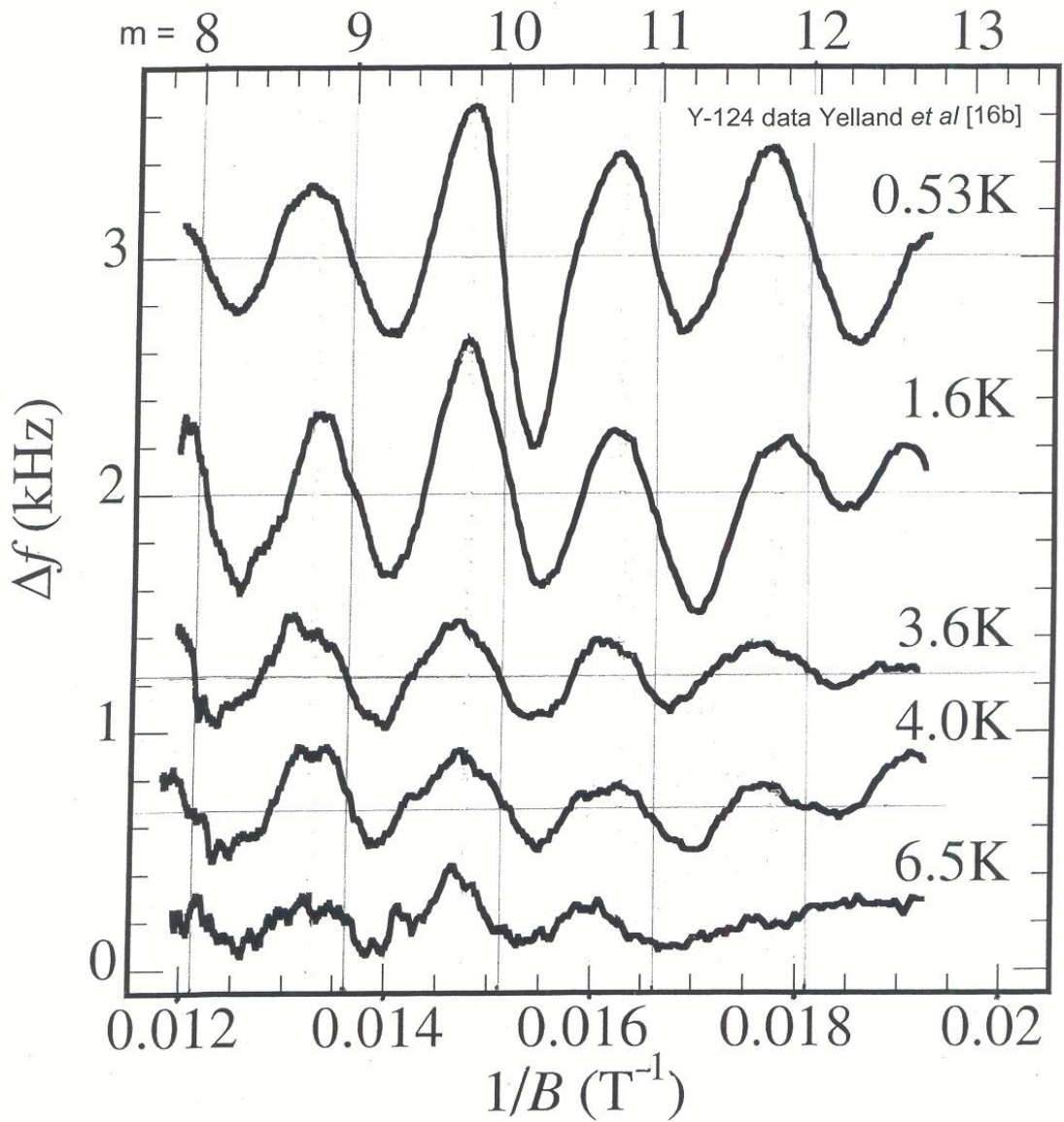

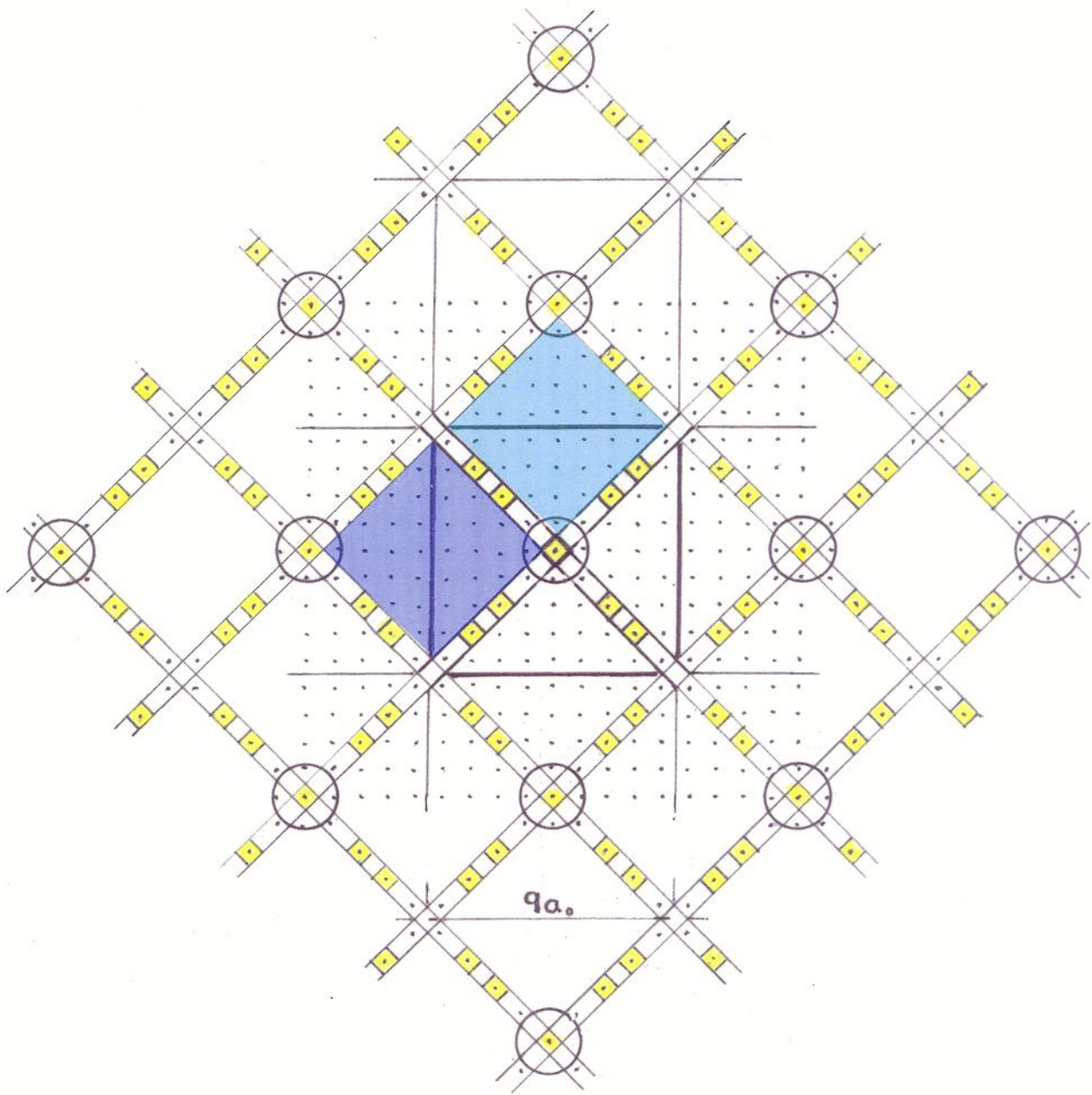

f17